\documentclass[11pt,a4paper]{article}
\pdfoutput=1
\usepackage[utf8]{inputenc}
\usepackage[english]{babel}
\usepackage{extarrows}
\usepackage{amsmath}
\usepackage{amsfonts}
\usepackage{amssymb}
\usepackage{graphicx}

\usepackage{caption}
\usepackage{subcaption}
\usepackage{float}
\usepackage{color}
\usepackage{cancel}
\usepackage{abstract}
\usepackage{appendix}
\usepackage{authblk}
\usepackage{xcolor}

\numberwithin{equation}{section}

\usepackage[hidelinks]{hyperref}

\newcommand\be{\begin{equation}}
\newcommand\ee{\end{equation}}
\newcommand\bea{\begin{align}}    
\newcommand\eea{\end{align}}      
\newcommand{\beb}{\begin{eqnarray}}
\newcommand{\eeb}{\end{eqnarray}}
\usepackage[top=2cm,bottom=2cm,left=2cm,right=2cm,marginparwidth=1.75cm]{geometry}

\title{{\bf Mutual Information of Generalized Free Fields}
\vspace{1.5cm}}

\author[]{Valentín Benedetti\footnote{valentin.benedetti@ib.edu.ar}}
\author[]{Horacio Casini\footnote{horacio.casini@cab.cnea.gov.ar}}
\author[]{Pedro J. Martinez\footnote{pedro.martinez@cab.cnea.gov.ar}}

\affil[]{\vspace{1cm} {\sl Centro At\'omico Bariloche and CONICET} \\
{\sl S. C. de Bariloche, R\'{\i}o Negro, R8402AGP, Argentina}}

\date{}
\usepackage[left=2cm,right=2cm,top=2cm,bottom=2cm]{geometry}

\begin{document}

\maketitle
\thispagestyle{empty}

\begin{abstract}
We study generalized free fields (GFF) from the point of view of information measures. We first review conformal GFF, their holographic representation, and the ambiguities in the assignation of algebras to regions that arise in these theories. Then we study the mutual information (MI) in several geometric configurations. The MI displays unusual features at the short distance limit: a leading volume term rather than an area term, and a logarithmic term in any dimensions rather than only for even dimensions as in ordinary CFT's. We find the dependence of some subleading terms on the conformal dimension $\Delta$ of the GFF. We study the long distance limit of the MI for regions with boundary in the null cone. The pinching limit of these surfaces show the GFF behaves as an interacting model from the MI point of view. The pinching exponents depend on the choice of algebra. The entanglement wedge algebra choice allows these models to ``fake''  causality, giving results consistent with its role in the description of large $N$ models.   
\end{abstract}

\newpage

\pagenumbering{arabic}
\tableofcontents

\section{Introduction}

Generalized free fields (GFF) are the simplest models of quantum field theories (QFT) satisfying Wightman's axioms \cite{greenberg1961generalized}. They are defined by having Gaussian correlations, that is, satisfying Wick's theorem for the $n$-point correlation functions. The theory is then completely specified by a two point function satisfying positivity, spectral condition, and Poincare covariance.  For a scalar field, the most general two point function  has the Kall\'en Lehmann form
\be
\langle \phi(x) \phi(y)\rangle= \int_0^\infty ds \, \rho(s)\, W_0(x-y,s)\,,
\label{GFF2}
\ee
with  $W_0(x-y,s)$ the two point function of a free scalar field of square mass $m^2=s\ge 0$. The spectral density $\rho(s)$ is a positive measure for $s\ge 0$ with at most polynomial increase in $s$.    

GFF appear naturally in some formal results in axiomatic QFT \cite{licht1965generalized,licht1961two,greenberg1962heisenberg,baumann1975field}. Due to the simplicity of the theory they have also been used in the mathematical literature as a source of examples to test different conjectures or analyze the independence or consistency of different properties, see for example \cite{yngvason1994note,Doplicher:1984zz,haag1962postulates}. From the physical point of view, they appear naturally as limits in large $N$ vector or matrix models \cite{coleman1988aspects}. The large $N$ limit suppresses higher truncated point functions with respect to the  two point functions for the symmetric fields. A notable example are holographic theories where generalized free fields describe the low energy sector of the theory in the large $N$ approximation, and are equivalent to ordinary free fields living in AdS space \cite{maldacena1999large,witten1998anti,dutsch2003generalized}.      

In this paper we study the entanglement entropy of GFF or, more precisely, we analyze the behavior of the mutual information in several cases of interest. Mutual information has the advantage of being regularization independent. However, the setup of the problem needs some distinctions to be made.

To expose the peculiarities that appear in entropic quantities for GFF as opposed to the case of more ordinary QFT let us consider a simple case first. When the GFF is a free field of mass $m$, the spectral density consists in a single delta function $\rho(s)=\delta(s-m^2)$.  In this case, an algebra of operators can be assigned to a spatial region $V$ at $x^0=0$. This algebra is generated by $\phi$ and $\dot{\phi}$ in $V$. Because of the hyperbolic equations of motion of the field, $(\square+ m^2) \phi=0$, this algebra coincides with the algebra generated by the field in the causal development of the spatial region $V$. In this case the entropy can be computed by the usual formulas for Gaussian states in terms of the field and momentum correlator at $x^0=0$, see for example \cite{Casini:2009sr}. 

If we now take $\rho(s)=\delta(s-m_1^2)+\delta(s-m_2^2)$, corresponding to the sum of two independent free fields $\phi(x)=\phi_1(x)+\phi_2(x)$, we could still apply the same formula in terms of correlators of $\phi$ and $\dot{\phi}$ at $x^0=0$. The algebra generated by $\phi,\dot{\phi}$ still closes in itself because of the numerical commutator of the GFF. However, notice that $\phi(x)$ now obeys an equation of motion with higher number of time derivatives $(\square+m_1^2)(\square+m_2^2)\phi=0$, so that $\ddot\phi$ and $\phi$ are independent operators. The inclusion or not of the operator $\ddot\phi$ leads to different algebras with different entropies. Considering just $\phi$ and $\dot{\phi}$ at $x^0=0$ will give us an entropy increasing like the volume of the region because it measures translational invariant local entanglement in field degree of freedom between $\phi=\phi_1+\phi_2$ and $\phi_1-\phi_2$. An analogous calculation can be found in \cite{chen2013quantum}.  If we now include $\ddot{\phi}, \dddot{\phi}$ in the algebra the result turns out to be exactly the algebra of two independent free fields of masses $m_1$ and $m_2$. This follows from
\be
\frac{\square+m_2^2}{m_2^2-m_1^2}\phi=\phi_1\;,\qquad\qquad \frac{\square+m_1^2}{m_1^2-m_2^2}\phi=\phi_2\;,
\ee 
from which we can reconstruct the two independent field and momentum operators.    
Hence, this new algebra containing higher derivatives of $\phi$ is equal to the algebra of the two fields $\phi_1,\phi_2$ in the causal development of $V$, and we get an area law rather than a volume law for the entropy. 

For a spectral density with any finite number $n$ of delta functions we have an analogous situation. We can take algebras of the field and less than $2 n-1$ time derivatives at $x^0=0$ and get a volume term for the entropy, or, provided we include $2 n-1$ time derivatives, the algebra and entropy will be the same as the one of $n$ independent free fields. In this last sense the $n$ independent free fields are encoded in a single GFF. 

In relativistic QFT it is natural to define the algebras taking a spacetime rather than a spatial region. If we take a finite time span around the spatial region $V$ there is no difference between the GFF defined with a finite number of delta functions in the spectral density and a theory of independent free fields. 
However, this discussion anticipates us the problems we can find when considering a continuous measure $\rho(s)$. In this case the theory has quite unusual properties. It does not satisfy the time slice axiom \cite{haag1962postulates}, meaning that the algebra generated by field operators in a finite time slice around $x^0=0$ does not exhaust all operators of the theory. This is another way to say that the field does not obey any local equation of motion, with any finite number of time derivatives. By the same reason it does not contain a stress tensor. Otherwise we could use it to construct the Hamiltonian in the algebra of a time slice. With the Hamiltonian we can then move operators in time to generate all operators in the theory. The Hamiltonian for a GFF with spectral measure having support in a non discrete set still exists but is rather non local. 

Then, it is clear that a spacial region does not determine uniquely an algebra for these theories and we must choose a spacetime region instead. A natural choice is to focus on causally complete regions. These are the domain of dependence of spatial regions.  However, even for a causally complete region there is in general an ambiguity in the algebra that can be associated to it for a GFF. Ambiguities on the assignation of algebras to regions appear also in ordinary QFT with generalized symmetries such as the Maxwell field \cite{casini2021entropic}, but they are much more severe for the GFF. 

A great simplification in the understanding of the nature of these ambiguities and the allowed algebras appear with the holographic realization of (a class) of these GFF as ordinary free fields in the bulk of a spacetime of one more dimension. 
We will focus on holographic GFF, specially in conformal GFF, and profit from the dual description to define the algebras for a given region and compute the entropy. 
In fact, independent computation of the entropies of the GFF in the boundary theory by standard methods without using the holographic description run into difficulties precisely because the nature of the algebras remain unspecified. For example, it is unclear how to apply the replica method because there is no action for the GFF. There is an action in the holographic bulk description that allows to apply the replica method there but the region/algebra in the bulk is not uniquely specified by the boundary region in the GFF. Large $N$ holographic theories choose automatically this bulk region through the gravity equations \cite{Lewkowycz:2013nqa}. Another boundary way of computing the entropy would be through formulas for Gaussian states in terms of correlation functions. This seems a complicated task. One should use correlation functions in the chosen space-time region using the methods of \cite{sorkin2014expressing}. This, however, has only access to one specific algebra for the region which is selected from the correlator. We will not attempt this calculation here.    
 
An outline of the contents of the paper is as follows. We first review GFF and their holographic description in the next section, and describe possible assignations of algebras to regions. Two choices of algebras are specially relevant. One of them, the causal wedge algebra, is more natural from the point of view of the GFF itself, while the other, the entanglement wedge algebra, is more relevant from the point of view of the limits of holographic large $N$ theories. 

In section \ref{finiteness} we show why the MI can be expected to be finite even is the AdS dual space is of infinite volume. 
In section \ref{short} we explore the short distance limit of the mutual information (MI). Interestingly, we find that the GFF has a volume law in this regime. This is in contrast to the case of ordinary theories where we have an area law in the short distance limit, even for theories coming from higher dimensions by Kaluza Klein dimensional reduction. The coefficient of the volume law can be computed and is universal in the sense that it does not depend on particular details of the GFF such as the conformal dimension. Other peculiarities include the existence of a logarithmic term for odd spacetime dimensions. We compute this logarithmic term in $d=3$ as a function of the GFF conformal dimension. 

In section \ref{long} we study the long distance limit of the MI. For spheres in CFT's this long distance limit is fixed by symmetry reasons and apply as well for conformal GFF \cite{Cardy:2013nua,Agon:2015ftl,Casini:2021raa}. For GFF the universality of this result can be understood by two reasons. The first is that there is a unique choice of algebras for spheres. The second is that spheres have a universal modular flow in CFT's. We show that the fact that general results for spheres apply to GFF imply certain holographic relations for the coefficients of the MI in general theories for different specific dimensions and spins whose reason would be rather mysterious otherwise. 
Taking non spherical regions, we study the case of regions with arbitrary boundaries in the light cone. This is useful to study the pinching limit of the MI in these theories by taking out a pencil of null generators from the null horizon of the region  \cite{Casini:2021raa,Agon:2021zvp}. Two particular pinching limits are relevant. One of them is a discriminator between free and interacting CFT. Free here is used in the sense of having a linear equation of motion rather than having Gaussian correlators. This gives us, as expected, vanishing MI in the pinching limit for all possible algebra choices of the GFF. The other limit is an indicator of violations of causality in the sense of the time slice axiom. For the causal wedge algebra we find causality violations while the entanglement wedge algebras avoids detection of causality violations in the GFF. This is a necessary condition for this algebras to come from the large $N$ limit of a theory with stress tensor. In both cases we compute the relevant pinching exponents for specific conformal dimensions. Here we make use of the results for the MI of free fields derived in \cite{Casini:2021raa}. We end with a discussion of the results.

\section{Conformal GFF and local algebras}
\label{GFFMI}
We first introduce conformal GFF fields. We will follow the description of \cite{dutsch2003generalized}. These have a spectral density given by a power law  
\be
\rho(s)= s^{\Delta-\frac{d}{2}}\,,\label{measure}
\ee
and conformal dimension $\Delta$. Eq. (\ref{measure}) gives a measure provided $\Delta$ obeys the unitarity bound $\Delta> (d-2)/2$.  For any such $\Delta$ the GFF defines a CFT. The case $\Delta=(d-2)/2$ is excluded because $\rho(s)$ becomes non integrable around $s=0$. The free massless field has $\rho(s)=\delta(s)$ instead. 

\begin{figure}[t]
\centering
\includegraphics[width=.68\linewidth]{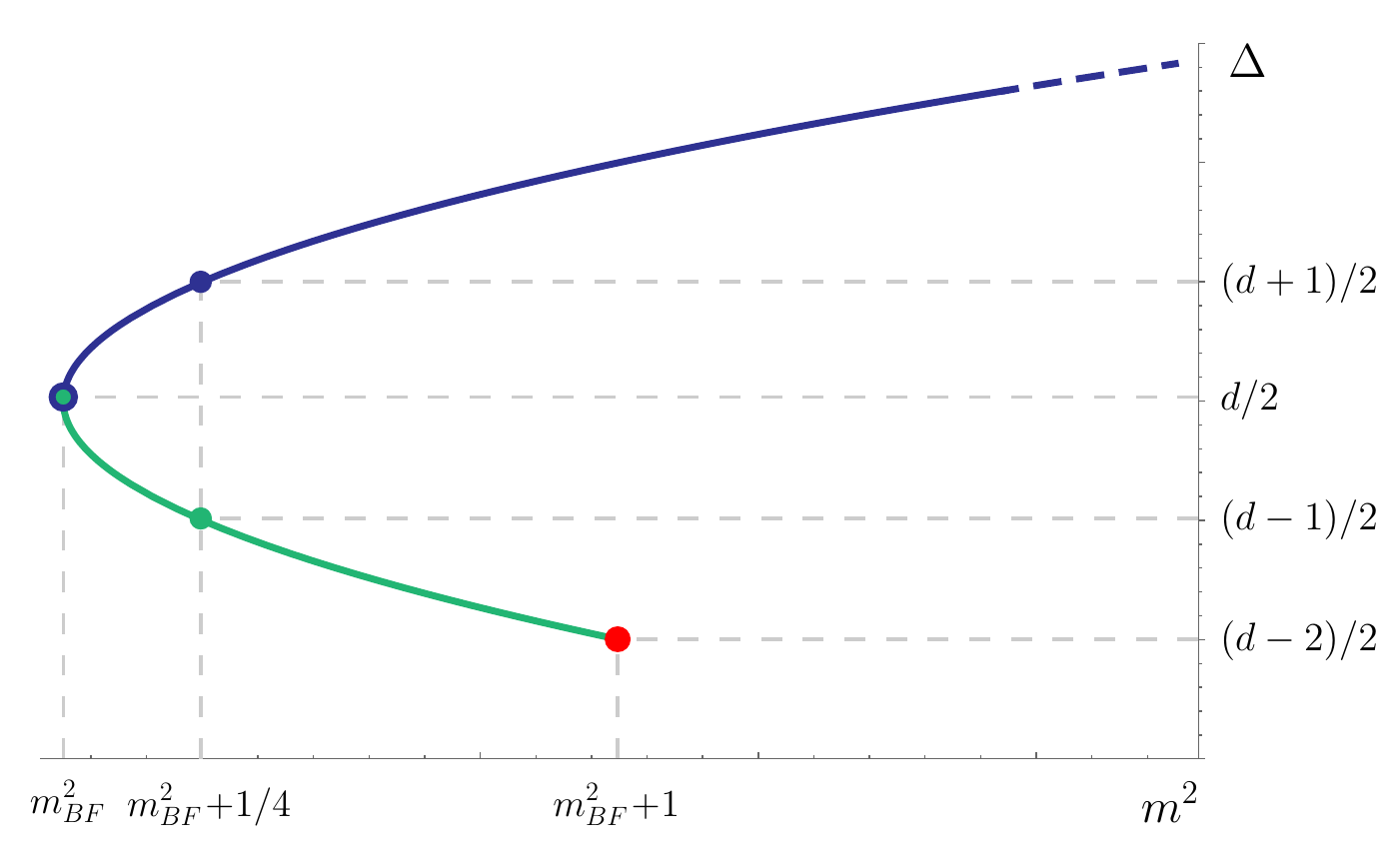}
\caption{\small The plot shows the relation \eqref{mass} and highlights some important points in the curve. The blue and green colored segments correspond to the standard and alternative quantization with Dirichlet and Neumann boundary conditions respectively. From bottom to top, the red dot is the end of the curve, where the CFT$_d$ reaches the unitarity bound, precisely at $m^2=m_{BF}^2+1$. The green dot shows the point where the massive AdS field is conformally coupled. The yellow and blue point is the BF mass bound, the lowest possible mass in AdS consistent with unitarity $m_{BF}^2=-d^2/4$. The blue point highlights the conformally coupled AdS field with the other boundary condition. The conformal dimension $\Delta$ at this point does not match the one of a free field in flat $d+1$ space.}
\label{Fig:BF}
\end{figure}

The holographic description is in AdS space. In the Poincare patch we write the metric
\be  
ds^2 = z^{-2} \, (dz^2+dx^2)\,,
\ee
with $dx^2$ the Minkoswki metric in $d$ spacetime dimensions and $z\in (0,\infty)$. The dual field $\varphi$ of the GFF is a free massive field in AdS with equation of motion 
\be
\left(z^2 \partial_z^2+z^2 \,\square_d+(1-d)z\partial_z -m^2\right)\varphi =0\,,
\ee
 where 
\be
m^2=\Delta (\Delta -d) \label{mass}
\ee
 can be negative. The minimal possible mass square is given by the Breitenlohner-Freedman bound  $m^2\ge m^2_{BF}=-d^2/4$ \cite{BFbound}.    
The field $\varphi$ can be canonically quantized with an AdS symmetric vacuum. For $-d^2/4\le m^2< -d^2/4+1$ there are are two inequivalent quantizations corresponding to the two roots of (\ref{mass}). These are defined by different boundary conditions for the field at the boundary $z=0$ of AdS.  Dirichlet boundary condition corresponds to $\Delta = \frac{1}{2}\left(d+\sqrt{d^2+4 m^2}\right)$ and Neumann boundary condition to $\Delta = \frac{1}{2}\left(d-\sqrt{d^2+4 m^2}\right)$.  See Fig. \ref{Fig:BF}. For $m^2\ge -d^2/4+1$ only the Dirichlet boundary condition is allowed. The limit $m^2\rightarrow -d^2/4+1$  of the Neumann branch hits the unitarity bound $\Delta\rightarrow (d-2)/2$. There is no holographic description of this point. There are also notable points at $m^2=-d^2/4+1/4$, $\Delta=(d\pm 1)/2$, in which the bulk is a massless conformally coupled scalar, and hence a conformal field. These particular bulk theories can be conformally mapped to half $d+1$-dimensional Minkowski space with metric $ds^2 =(dz^2+dx^2)$, where we have the two possible conformal boundary conditions at $z=0$. The Neumann branch at this point has $\Delta =(d-1)/2$ corresponding to a free massless $d+1$ dimensional free field, and the Dirichlet branch has a different boundary dimension $\Delta=(d+1)/2$ due to the boundary condition.

The relation between the boundary and bulk fields can be described as follows. The GFF is the boundary limit of the bulk field, 
\be
\lim_{z\rightarrow 0} z^{-\Delta} \, \varphi(x,z)=\frac{2^{-\alpha-1/2}}{\Gamma[\alpha+1]}\, \phi_\Delta(x)\,,\label{li}
\ee  
where $\alpha=\Delta-d/2$, while the bulk field has a non local expression in terms of the boundary one
\be
\varphi(x,z)=\frac{1}{\sqrt{2}}\, z^\Delta \, (z^2 \square)^{-\alpha/2}\, J_\alpha(z \sqrt{\square})\, \phi_\Delta(x)\,.
\ee 
 
\subsection{Algebras}

A more illuminating relation between bulk and boundary theories is given in terms of local algebras. If $W$ is a region in AdS let us call $W'$ to the set of points spatially separated from $W$ in the bulk. The causal completion of $W$ is $W''$ and a causally complete region satisfies $W=W''$. Causally complete regions in the bulk are the domain of dependence of spatial surfaces and are naturally attached to algebras ${\cal A}_\varphi(W)$ generated by the bulk free field $\varphi$ in $W$. 

In the boundary theory, for any space-time region $U$ let us call ${\cal A}_\phi (U)$ to the algebra generated by the GFF $\phi$ in $U$. If we consider $U$ as a region in the boundary of AdS we can define an associated causal region in the bulk as $U''$. For a double cone $D$  (the intersection of the past of a point with the future of another point) in the boundary it was shown in \cite{dutsch2003generalized} that we have the equality
\be
{\cal A}_\phi(D)={\cal A}_\varphi(D'')\,. \label{gece} 
\ee

This relation can be generalized. The boundary algebras ${\cal A}_\phi (U)$ are generated by the local GFF and then are additive under union of spacetime regions, that is, we can decompose them as generated by double cone algebras
\be
{\cal A}_\phi (U)=\bigvee_{D\subset U} {\cal A}_\phi(D)=\bigvee_{D\subset U} {\cal A}_\varphi(D'') \,.\label{pp0}
\ee
In particular, if $U$ is causally closed in the boundary, we have
\be
{\cal A}_\phi (U)={\cal A}_\varphi(J^+(U)\cap J^-(U)) \,,\label{pp1}
\ee
where $J^+(U)$ and $J^-(U)$ are the bulk future and past of $U$. This generalizes (\ref{gece}). The 
bulk region $U_{CW}=J^+(U)\cap J^-(U)$ is called the causal wedge of $U$  \cite{hubeny2012causal}.\footnote{This region is not in general a bulk causally closed region. Therefore, a natural expectation is that using properties of the algebras of free fields \cite{araki1964neumann}, the bulk region in the left hand side of (\ref{pp1}) could be extended to $((J^+(U)\cap J^-(U))''=U''$, having same algebra. We will not need this in the following.}   Therefore, smearing the GFF for such $U$ one obtains the bulk free field algebra in the causal wedge.  This assignation of algebra is the most natural one from the point of view of the GFF and is also the minimal possible one, being generated by the GFF in the region. It will be called the causal wedge algebra. We write
\be
{\cal A}_{CW}(U)={\cal A}_\phi(U)={\cal A}_\varphi(U_{CW}).
\ee

\begin{figure}[t]
\centering
\includegraphics[width=.5\linewidth]{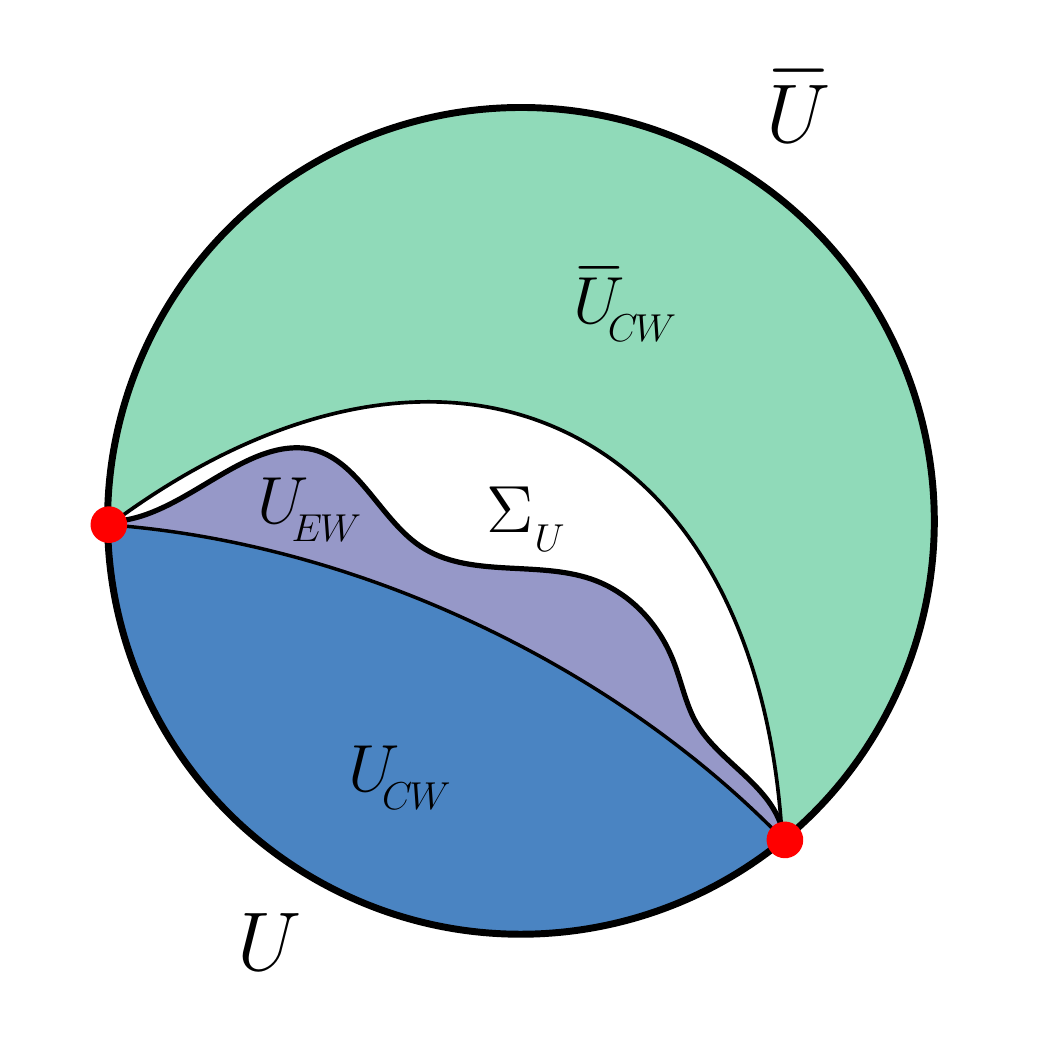}
\caption{\small A constant time cut of AdS. The causal wedges $U_{CW}$ and $\bar{U}_{CW}$ for complementary boundary regions $U$ and $\bar{U}$ do not exhaust the space. A surface dividing the causal shadow in two, can be used to define complementary algebras. The HRT surface $\Sigma_U$ gives one such partition defining the entanglement wedge $U_{EW}$. }
\label{Fig:Wedges}
\end{figure}

We can also define the causal complement in the boundary spacetime as $\bar{U}$, and a causally complete region in the boundary satisfies $U=\bar{\bar{U}}$.\footnote{We take the complement $\bar{U}$ in the compactified space or equivalently in the spacetime cylinder.}  By causality, algebras corresponding to complementary boundary regions commute:
\be
 {\cal A}_\phi (U)\subseteq ({\cal A}_\phi (\bar{U}))',
\ee
where ${\cal A}'$ is the commutant of ${\cal A}$, that is, the set of operators that commute with those of ${\cal A}$. For a general QFT when there is equality ${\cal A}(U)= ({\cal A}(\bar{U}))'$ the theory is said to satisfy Haag duality for $U$. An ordinary free scalar field satisfies duality for any causal region \cite{araki1964neumann}. Then, from the bulk representation (\ref{pp1}) we can easily check that the causal wedge algebras of the GFF do not satisfy duality for general regions. This is because to the boundary complementary regions $U$ and $\bar{U}$ it corresponds the bulk regions $U_{CW}$ and $\bar{U}_{CW}$ which generally fail to be complementary. The region spacelike to these two regions $(U_{CW}\cup \bar{U}_{CW})'$ is called the causal shadow \cite{headrick2014causality}. See Fig.\ref{Fig:Wedges}. An exception is the case where $U$ is a double cone and the causal shadow vanish. Haag duality for double cones is in fact always necessarily valid for all CFT's, where the complement is taken in the compactified space \cite{brunetti1993modular}.    

Examples of failure of Haag duality are also known for more familiar theories obeying the time slice axiom. In these examples the problems in the relations of algebras and regions are related to the existence of operators in regions with non trivial topology which cannot be generated locally by field operators in the same region. These topological failures of Haag duality are associated to generalized symmetries  and are absent for sufficiently complete theories \cite{casini2021entropic,casini2021completeness}.  A simple example is the case of the free Maxwell field which does not satisfy Haag duality for regions with the topology of a ring due to the existence of Wilson loop operators. However, the present case is different in several aspects. The failure of duality for GFF is related to the failure of the time slice axiom and the consequent failure of additivity for causal regions based on the same spacial plane. The algebra generated by the field in two overlapping double cones does not correspond to the algebra of a causal region for a GFF. In contrast, ordinary QFT examples satisfy this form of causal additivity. Moreover, the relative commutant ${\cal A}(U)' \cap {\cal A}(\bar{U})'$ is trivial in ordinary QFT examples while it is a large algebra  for the GFF. This  is represented by the algebra of the bulk fields in the causal shadow region, see Fig.\ref{Fig:Wedges}. 

Once Haag duality fails, the possible assignation of algebras to regions is not unique. We can enlarge the algebras of $U$ and $\bar{U}$ by keeping then still commuting to each other. In the holographic representation, a simple way to do this is by moving the boundaries of the the associated algebra of bulk field towards the bulk but keeping them spatial to each other. If we partition the bulk in two regions, one containing $U_{CW}$ and the other $\bar{U}_{CW}$, the associated free field algebras will be dual to each other, and we can recover Haag duality. This is a possible Haag-Dirac net as defined in \cite{casini2021entropic}. The prescription should also be monotonically increasing with the region size to give larger algebras to larger regions. 

This is precisely what the holographic prescription does. It selects a particular division of the space in two given by the minimal surface $\Sigma_U$ anchored at the boundary of $U$ (or equivalently $\bar{\bar{U}}$). This is called the RT \cite{Ryu:2006bv} or HRT  \cite{Hubeny:2007xt} surface. Here $U$ is assumed to be a boundary causal region. The bulk causal region spanning from  $U$ to $\Sigma_U$ is called the entanglement wedge $U_{EW}$. It follows that $U_{CW}\subseteq U_{EW}$, and that the mapping $U\rightarrow U_{EW}$ is monotonic under the inclusion order. This property is called entanglement wedge nesting \cite{Wall:2012uf}.  The corresponding assignation of algebras to the boundary regions will be called the entanglement wedge algebra 
\be
{\cal A}_{EW}(U)={\cal A}_\varphi (U_{EW})\,.
\ee
This is in fact the algebra of low dimension operators attached to the region in the large $N$ limit of holographic models. 
By construction it 
follows that 
\be
{\cal A}_{CW}(U)\subseteq {\cal A}_{EW}(U)\,,\hspace{.7cm}  {\cal A}_{EW}(U)=({\cal A}_{EW}(\bar{U}))'\,.
\ee
 
We remark that from the point of view of the GFF theory itself there are potentially infinitely many different choices of algebras for the regions that satisfy duality and the nesting property and the entanglement wedge is just one of them. An important open question is whether there is some intrinsic GFF argument that selects the entanglement wedge as a preferred choice. For holographic theories there is the idea of bulk reconstruction, i.e., reconstruction of bulk operators in $U_{EW}$ from the boundary operators in $U$. At the level of the GFF this reconstruction is done using the modular flow of ${\cal A}_{EW}(U)$ \cite{faulkner2017bulk}. But this modular flow already involves the bulk operators in the region. In principle, for other bulk regions different from the entanglement wedge, one could reconstruct the bulk fields in a like manner.   

\subsection{Mutual information}

Coming to the problem of computing the mutual information for a GFF, we see that there is no unique definition of the MI between two regions. Namely, one should also choose the specific algebras one is assigning to them. To define a precise problem we turn to a holographic description. A choice of algebra can then be made via a choice of bulk region and the unique algebras of a free field assigned to it.

The comparison with the large $N$ limit of a holographic theory is as follows. This large $N$ limit selects the entanglement wedge, a privileged bulk region which is bounded by the RT surface. The bulk free field on this region represents the low dimension CFT algebra coming from a UV complete theory and thus avoids all possible ambiguities in the region-algebra correspondence. One possible way to define the mutual information for a GFF assigns precisely the entanglement wedge algebras: 
\be\label{III}
I_{EW}(A,B)=I_\varphi(A_{EW},B_{EW}).
\ee
However, while this quantity is always sensible for the theory of the GFF, it is not always relevant for the large $N$ theory itself. In this later theory MI admits a decomposition in terms of heavy and light operators as follows. We split between $\Delta\gg c$ and $\Delta\ll c$ operators \cite{Hartman:2014oaa}
\begin{equation}
I_{CFT}(A,B)\sim I^{\Delta\gtrsim c}_{CFT}(A,B)+ I^{\Delta \ll c}_{CFT}(A,B)\;.
\end{equation}
The rhs of \eqref{III} also splits in terms of a leading $G^{-1}\sim N^{2}$ contribution coming from the RT area term and a $G^{0}\sim N^{0}$ contribution coming from the free fields living inside the causal wedges  \cite{Faulkner:2013ana}, i.e.
\begin{eqnarray}
I^{\Delta\gtrsim c}_{CFT}(A,B) &=& \frac{1}{4G}\left( A(\Sigma_{A})+A(\Sigma_{B})-A(\Sigma_{A\cup B}) \right)\,,\\
  I^{\Delta \ll c}_{CFT}(A,B) &=& \sum_\varphi (S_\varphi(A_{EW})+S_\varphi(B_{EW})-S_\varphi((A\cup B)_{EW}))\;.
\label{RTMI}
\end{eqnarray}
The combination of area terms describes the $\Delta\gtrsim c$ operator's contribution to MI, while the last term gives the bulk contribution by the free fields $\varphi$ representing the sector of $\Delta\ll c$ in the CFT. To leading order in $N^{-1}$, the bulk fields $\varphi$ are free and the contribution to the MI of each eventual conformal dimension $\Delta\ll c$ decouple on the rhs. Two opposite situations may occur in applying this formula. If $A$ and $B$ are far enough the RT surface of the union decouple into the union of RT surfaces $\Sigma_{A\cup B}=\Sigma_{A}\cup \Sigma_{B}$. In this case the area term cancel in the MI and we get
\be
 I_{CFT}(A,B)\sim S_\varphi(A_{EW})+S_\varphi(B_{EW})-S_\varphi(A_{EW}\cup B_{EW})=I_{EW}(A,B)\,.
\ee
The CFT MI coincides with the causal wedge MI of the GFF to leading order in $N$. On the other hand, when $A$ and $B$ are close enough, there is a phase transition to a connected RT surface such that $\Sigma_{A\cup B}\neq\Sigma_{A}\cup \Sigma_{B}$. In this case not only the area term in the MI does not vanish but the subleading piece (rhs of (\ref{RTMI})) is not the mutual information between any regions in the bulk. In fact, it is a combination of bulk entropies whose boundaries do not match, and hence it is not free from UV ambiguities and divergences from the point of view of the bulk free field. In this case, the order $N^0$ term cannot make sense without the presence of the area term. Indeed, it is expected that the divergences and ambiguities of the order $N^0$ entropies to be renormalized in the $G^{-1}$ coefficient of the area term. However, to our knowledge this calculation has not been made precise in the literature yet. In any case, it is clear that after the phase transition in the RT surface the mutual information of the CFT looses contact with the GFF MI, which cannot be considered even a subleading contribution. As we will see, this is necessary because the GFF MI at short distances increases beyond what is expected for an ordinary CFT.  
 
In the following sections we analyze the MI of the GFF theory itself and we will explore both short and long distance leading terms. 
  
\section{Finiteness of the MI}
\label{finiteness}
 In the bulk there is an ordinary massive free field but it lives in a hyperbolic space-time. In the near boundary region the distance becomes large and correlations should decay but the volume of the regions also increase to infinity. 
This rises the question of whether the contribution of the near boundary region to the MI remains finite for different $\Delta$. For the conformally coupled bulk the warp factor can be eliminated and the full MI is clearly finite.  In this case, whilst correlators fall off near the boundary the area increases  such as to compensate the decay of correlation, but for generic $\Delta$ the situation is unclear. For example, in the Neumann branch the correlator behaves as $G_N\sim(\frac{z}{\epsilon})^{2 \Delta}$ which decays more slowly than the increase in the area $\sim z^{-(d-1)}$ for $\Delta <(d-1)/2$, below the conformally coupled point. The intuition may lead us to think that this regime could lead to divergences in the MI. We will see below that this is not the case.

\begin{figure}[t]
\centering
\includegraphics[width=.45\linewidth]{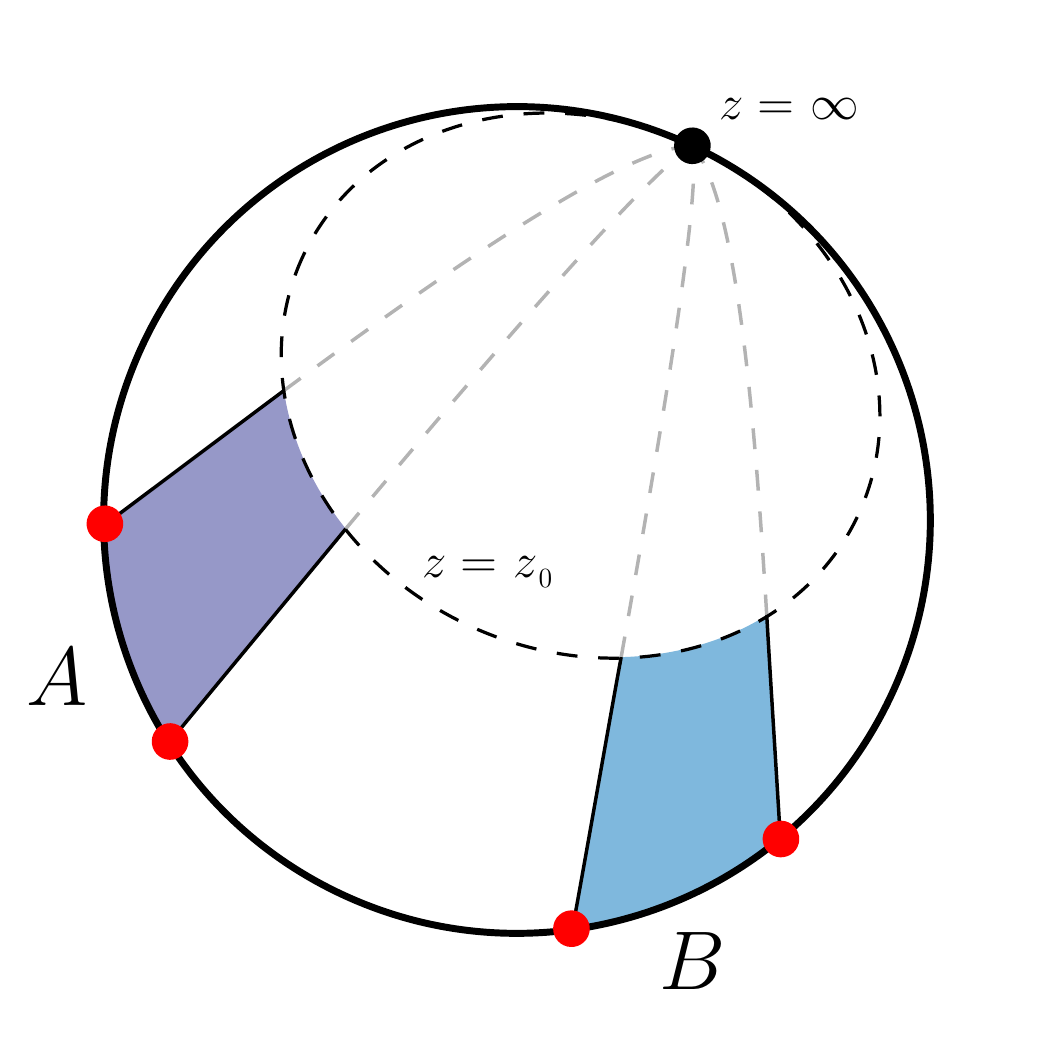}
\caption{\small Two regions in AdS that are straight in the $z$ direction and end at $z=z_0$. Their asymptotic boundaries define subregions $A$ and $B$ in the CFT.}
\label{Fig:wedges}
\end{figure}

In order to settle this point we make the following computation. We need to estimate the contribution of the MI coming from the region close to the boundary between two systems. Then we can simplify our calculation to the one of the MI between two straight walls in AdS from $z=0$ to a irrelevant infrared cutoff placed at a fix but arbitrary $z=z_0$. Hence the regions have topology $A\times I$, $B\times I$, were $I$ is the interval $z\in (0,z_0)$. See Fig.\ref{Fig:wedges}. Since we are interested in investigating the finiteness of the near boundary contribution to the MI we can slightly change the model and impose a boundary condition $\Phi(x,z=z_0)=0$ at the infrared cutoff $z=z_0$. This boundary condition just simplifies the evaluation of the MI.  

Given this geometric configuration we reduce the AdS massive scalar field action to a tower of massive scalars in flat $d$ dimensional spacetime given by the boundary coordinates $x$. 
We start from the AdS$_{d+1}$ action for the massive scalar in Poincarè AdS
\be
S=\frac{1}{2}\int dz\, d^dx \, \sqrt{g}\, \varphi \left(  \square  - m^2 \right)\varphi+ b.t.  \label{Sactionads2}
\ee
where $b.t.$ stand for boundary terms that are not relevant in our analysis. The action is defined over $z\in (0,z_0)$ at which end-points one should impose either Dirichlet or Neumann boundary conditions. We impose $\varphi(x,z=z_0)=0$ and the boundary conditions at $z=0$ is fixed either to the standard or alternative quantization, giving $\Delta=d/2\pm\sqrt{d^2 /4+m^2}$ respectively. The equations of motion are
\begin{equation} 
     \left(  \square  - m^2 \right)\varphi=\left(z^2 \partial_z^2+z^2 \,\square_d+(1-d)z\partial_z -m^2\right)\varphi=0\;.
\end{equation} 
We now expand the field as
\begin{equation}
    \varphi(x,z)=\sum_{\kappa}  z^{d/2} J_{\alpha} (\kappa z) \varphi_\kappa(x)\;,
\end{equation}
where $\alpha=\Delta-d/2$. 
 This has the correct boundary condition\footnote{To be explicit, $\alpha\geq0$ describes the standard quantization whilst $-1<\alpha<0$ describes the alternate quantization and $\alpha=-1$ sits at the AdS unitarity bound, cf. Fig. \ref{Fig:BF}.} at $z=0$ for any $\Delta$. The condition at $z=z_0$ is fulfilled provided the values of $\kappa$ are quantized as the zeroes of the Bessel function
\be
J_{\alpha}(\kappa\, z_0)=0\,.\label{esr}
\ee 
In this basis, we get
\begin{align}
S & = \frac{1}{2} \sum_{\kappa,\kappa'}\, \int  d^dx \, \varphi_{\kappa'}(x) \left(  \square_d - \kappa^2 \right)\varphi_\kappa(x) \left( \int_0^{z_0}  dz \, z J_{\alpha} (\kappa z) J_{\alpha} (\kappa' z)\right) + b.t. \nonumber \\
& = \sum_{\kappa}\,\left(\frac{z_0\,J'_{\alpha} (\kappa z_0)}{2}\right)^2 \int d^dx \, \varphi_{\kappa}(x) \left(  \square_d - \kappa^2 \right)\varphi_\kappa(x) + b.t. \,. 
\end{align}
The first line is finite and the orthogonality relations for $J_{\alpha}$ was used. We conclude that $\kappa$ plays the role of the mass of the $d$-dimensional fields. 

At this point, we have shown that the system we are describing has the same dynamics as tower of scalar fields in a $d$-dimensional flat spacetime, with masses $\kappa_n$, $n=1,2,\cdots$, given by the zeros of the Bessel function (\ref{esr}), forming a discrete spectrum. By our choice of region the algebras of the bulk field is just the tensor product of the algebras of the two dimensional modes. Then the entropies and MI are additive for each mode \cite{Casini:2009sr}
\be
I_\varphi(A\times I,B\times I)=\sum_n I_0(\kappa_n,A,B)\,,\label{su}
\ee
where $I_0(\kappa,A,B)$ is the MI of a $d$-dimensional flat space scalar with mass $\kappa$.
Since we are now dealing with standard non divergent MI in flat space, the only possible divergences for any AdS set-up must be apparent in the $\kappa$ spectrum given by eq. (\ref{esr}). For large $n$ we have $\kappa_n\sim (\pi/z_0)\, n$.  Then, the sum of the MI of the two dimensional modes is finite since the MI $I_0(\kappa,A,B)$ decays exponentially for large mass \cite{casini2005entanglement}.   

The limit of the unitarity bound $\Delta\rightarrow (d-2)/2$ corresponds to $\alpha\rightarrow -1$. In this limit the first mode has mass
\be
\kappa_1\sim \frac{2}{z_0}\, \sqrt{\Delta-\frac{d-2}{2}}\to 0\,.\label{k1}
\ee 
For $d>2$ this is in accordance with the fact that the GFF MI must converge to the one of a massless free field in this limit.  For $d=2$ the massless free field is not a well defined theory since it has an IR divergent correlator. This gives place to an IR divergent term for $d=2$ and small $\kappa$ \cite{Casini:2009sr}
\be
I_0^\kappa\sim  \frac{1}{2}\log(-\log(\kappa R))\,,\hspace{.7cm}\kappa R\ll 1\,,\,\,\,\, d=2\,,\label{ll}
\ee
where $R$ is fixed by the geometry of $A,B$. This gives an IR divergence in the MI as a function of $\Delta$ in the unitarity bound limit,
\be
I_{GFF}\sim \frac{1}{2}\,\log\left(-\log\left(\Delta\right)\right)\,,\hspace{1cm} \Delta\ll 1\,,\,\,\,\, d=2\,.
\ee
 This divergence is independent of the geometry and happen for any other pairs of bulk regions attached to the boundary, in particular for the all possible algebras of the GFF.

Another commentary is that if we take the limit $z_0\rightarrow \infty$ in (\ref{su}) the mass spectrum becomes dense and we can replace the sum by an integral
\be 
I_\varphi(A\times I,B\times I)\sim \frac{z_0}{\pi}\,\int_0^\infty d\kappa\, I_0(\kappa,A,B)\,.
\ee
The integral converges but the mutual information diverges as $z_0$. This divergence is natural since the bulk regions touch each other in the point $z\rightarrow \infty$ in this case, see Fig.\ref{Fig:wedges}. The bulk algebras of the form $A\times {z\in (0,\infty)}$ form another set of algebras that can be attached to boundary regions in the GFF. This prescription is Poincare covariant but not conformal invariant. This algebra assignation is called the dual net in the mathematical literature because it arises as the commutant of the algebra of the complement of the region inside Minkowski space (rather than the spacetime cylinder). It was shown that this algebra assignation does not satisfies the split property \cite{Doplicher:1984zz}, preventing the calculation of a meaningful MI. This is consistent with the present  results.

\section{Short distance mutual information \label{short}}
\label{GFFfromAdS}

In this section, we compute the MI of the GFF in the limit of short distances. For simplicity, we choose to analyze the simplest case of a sphere $A$ of radius $R^-$ and the complementary region $B$ of the
sphere of radius $R^+$. For conformal GFF there is no issue of different possible algebra choices for this geometry. 
We assume the regions to be close together. This is $0<R^+-R^-=\epsilon\ll R^{\pm}$. Since our theory is a CFT, only dimensionless quotients of $R=(R^++R^-)/2$ and $\epsilon$ can appear.
We are thus set to compute the MI $I_{GFF}(A,B)$ via holography,
\begin{equation}
I_{GFF}(A,B)=I_\varphi(A_{CW},B_{CW})\;.
\label{II}
\end{equation}
We have that $A_{CW}$ and $B_{CW}$ are bulk hemi-spheres of radii $R^{\pm}$ in the bulk standard coordinates $x,z$. At the end we comment on the modifications that arise for non spherical regions. 

\subsection{The conformal bulk case $\Delta=(d\pm 1)/2$}\label{SecConfBulk}

The analysis is simpler if we begin by considering the $m^2=(1-d^2)/4$, corresponding to two possible conformal dimensions $\Delta=(d\pm 1)/2$. In this case the bulk field is conformally coupled and the theory can be mapped via a Weyl transformation $ds^2=z^{-2} (dz^2+dx^2)\rightarrow ds^2=(dz^2+dx^2)$ to a massless free field in $d+1$ flat space with a wall.
The MI is invariant under the Weyl transformations and we end up with a problem of a free massless field in flat space. 
The leading contribution comes from local entanglement along the two nearby boundaries of the bulk semi-spheres which are at a fixed distance $\epsilon$ between each other. It is an extensive contribution along the boundary and then will be proportional to the area of the semi-sphere
\begin{equation}\label{MILocalaprox}
 I_{GFF}\sim k_{d+1} \,\int \frac{dA}{\epsilon^{d-1}}=k_{d+1} \, \frac{\pi^{d/2}}{\Gamma(d/2)}\,\frac{R^{d-1}}{\epsilon^{d-1}}\,.
\end{equation}
The constant $k_d$ is the coefficient in the area term in the mutual information for a free massless scalar between two planar boundaries in $d$ dimensions. It can be computed in terms of solutions of a Painleve equation by dimensional reduction to a $d=2$ massive field problem \cite{casini2005entanglement,Casini:2009sr}.\footnote{We have, for example, $k_3=0.0396506498...$, $k_4=0.0055351600...$, $k_5=0.0013139220...$ . For higher dimensions a good approximation is 
$
k_d\simeq \frac{\Gamma\left(\frac{d-2}{2}\right)}{2^{d+2}\, \pi^{\frac{d-2}{2}}}
$.}
The surprising feature of (\ref{MILocalaprox}) is that the MI of the GFF has a {\sl volume law} rather than the area law that holds for ordinary QFT. The coefficient of the volume law is the same for the two boundary conditions. A similar ``wrong dimensionality'' formula is found for the thermal partition function of GFF  \cite{el2012emergent}.

Though the general form if the MI as a function of $R/\epsilon$ is complicated, some subleading terms in the limit of large $R/\epsilon$ also follow from known results for ordinary free fields. This limit of the mutual information can be understood as a regularization of the entropy of a bulk semi-sphere in presence of the wall boundary conditions, where $\epsilon$ plays the role of the cutoff and we have to identify $I\sim 2 S$ \cite{Casini:2007dk}. Hence, subleading terms in the mutual information follow the same pattern as subleading terms in the entropy, and universal terms in the entropy can be directly related to terms in the mutual information. These terms arise either from subleading bulk contributions or from boundary contributions. For example, for $d=2$ ($d+1=3$ bulk) there is no bulk logarithmic term but there is a boundary logarithmic term that depends on the particular conformal boundary condition, induced by the so called $b$ anomaly coefficient \cite{kobayashi2019towards,fursaev2016anomalies,jensen2019weyl}. For a free scalar with Dirichlet and Neumann boundary conditions these are computed in \cite{solodukhin2016boundary,fursaev2016anomalies}, see also \cite{nozaki2012central,jensen2016constraint}. We have in this case
\beb
I_{GFF} & =  & \pi\, k_3\, \frac{R}{\epsilon} - \frac{1}{6} \log\left(\frac{R}{\epsilon}\right) +\cdots\hspace{1cm}  d=2\,, \hspace{.3cm}\Delta=\frac{d+1}{2}=\frac{3}{2}\,, \\  
I_{GFF} & = & \pi\, k_3\, \frac{R}{\epsilon} + \frac{1}{6} \log\left(\frac{R}{\epsilon}\right)   +\cdots \hspace{1cm}d=2\,, \hspace{.3cm}\Delta= \frac{d-1}{2}=\frac{1}{2}\,.
\eeb

A logarithmic term appears in $I_{GFF}$ in any dimensions. For even $d$ this logarithmic term is induced by the boundary conditions and depends on $\Delta$ as in the previous example, 
\be
I_{GFF}^{\rm log} = (-)^{d/2}\, 2\, B' \, \log(R/\epsilon)\,,\hspace{1cm} d \,\,\,\textrm{even}\,, 
\ee
with $B'$ a coefficient for a scalar depending on the chosen boundary conditions \cite{kobayashi2019towards}. 
For odd $d$ is induced by the spherical surface of the $d+1$ dimensional bulk and is independent on $\Delta$. It is proportional to the usual logarithmic contribution for the entropy of a whole sphere since we have a half sphere but the mutual information multiplies contributions by $2$. Thus we have 
\be
I_{GFF}^{\rm log}= (-)^{(d-1)/2}\,2\, A_{d+1} \, \log(R/\epsilon)\,,\hspace{1cm} d \,\,\,\textrm{odd}\,, \label{odd}
\ee
where $A_{d+1}$ is the trace anomaly coefficient of a free scalar in $d+1$ dimensions. 

\subsection{Volume term for any $\Delta$}    
When the bulk field is not conformally coupled ($m^2\neq (1-d^2)/4$) we cannot conformally transform to flat space. The bulk free field MI is one  of a massive field in AdS with boundary conditions. However, the short distance leading term in the MI can still be computed in a similar manner. 

Suppose we have a free massless scalar in flat $d+1$ dimensional Minkowski space. The leading contribution to the MI between two nearby entanglement surfaces separated by a distance $l(x)$ is given by 
\be
I\sim k_{d+1} \int \frac{dA}{l(x)^{d-1}}\,.\label{basically}
\ee
This is a local area term produced entirely from local correlations of nearby operators across the boundaries. In this formula it is assumed that the distance changes slowly in the scale of the distance itself $|\nabla l(x)|\ll 1$. The presence of distant boundary conditions cannot change this local contribution. For a massive field the same formula applies if the distance scale is smaller than the mass scale $ m\, l(x)\ll 1$. When $m\, l(x)\gtrsim 1$ the correlations across the gap between the regions fall exponentially and the contribution is cutoff. For a curved space the same formula applies if the distance scale is much smaller than the curvature scale.   

\begin{figure}[t]
\centering
\includegraphics[width=.55\linewidth]{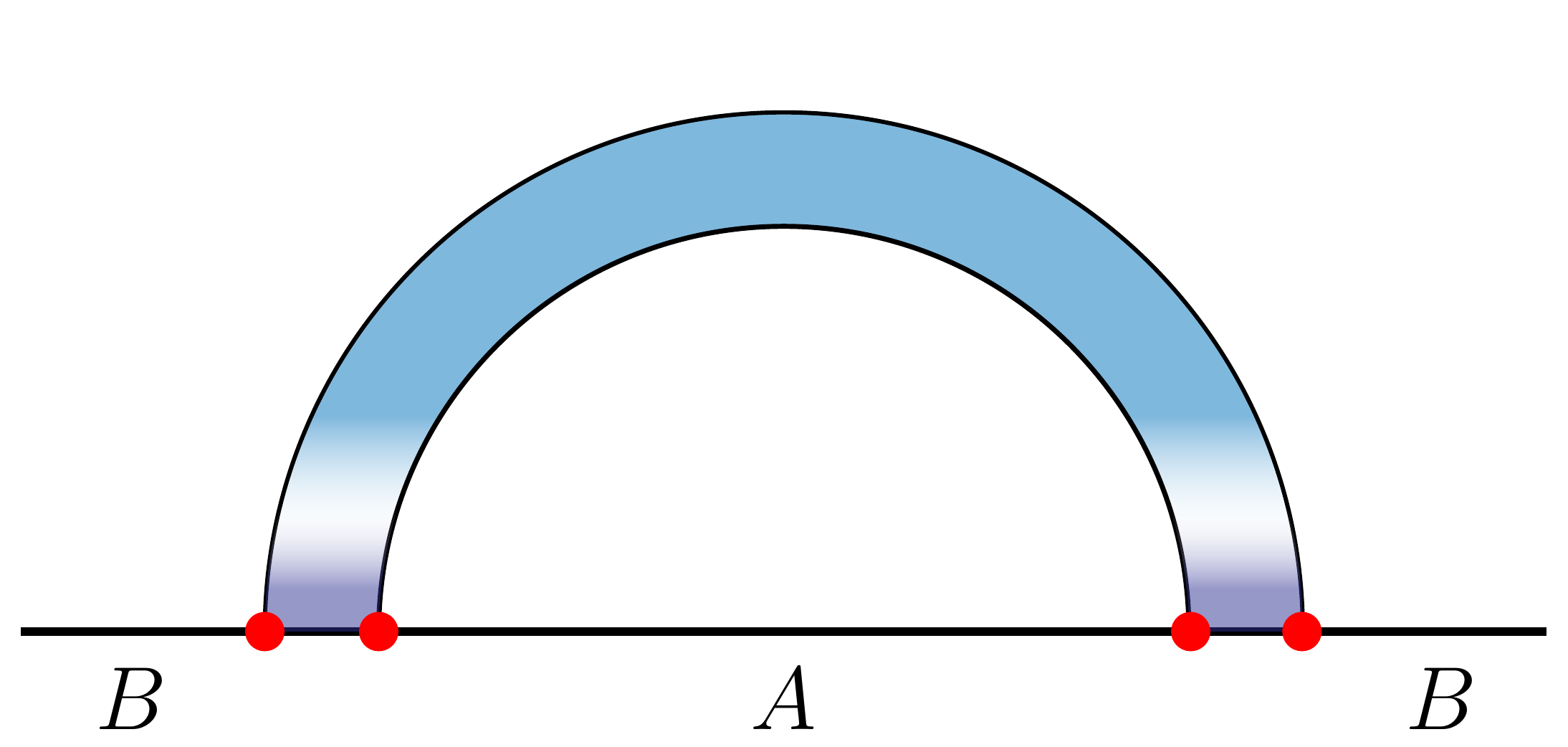}
\caption{\small In the light-blue region between the semi-spheres, $z\gg \epsilon$ and $z\gg \epsilon \Delta $, mass and curvature scale can be neglected to compute the short distance bulk entanglement contribution. The rest of the contribution, in violet, can be thought as a boundary term.}
\label{Fig:SDMI}
\end{figure}

In the present situation the distance between the two boundaries in the metric $ds^2=(dx^2+dz^2)$ is still $\epsilon$, and therefore, for small $\epsilon$, the physical distance is $l(z)=\epsilon/z$, getting smaller deeper into the bulk. We have $|\nabla l(z)|= \epsilon/z$. The curvature scale of the AdS was set to $1$. Hence, in both cases the condition for the application of the formula for the area term is $z\gg \epsilon$.  If the field mass is much larger than one we also need that the distance is smaller than the inverse mass $m \,\epsilon/z \ll 1$.  See Fig.\ref{Fig:SDMI}. Thus we get
\be
I_{GFF} \sim k_{d+1}\,\int_{z\gg \epsilon, \epsilon \,m} \frac{dA}{(\epsilon/z)^{d-1}}\,.     
\ee
This is integrated on the surface $z^2+r^2=R^2$. This is again the same integral (\ref{MILocalaprox}) once we take into account the area element is scaled by $z^{-(d-1)}$ with respect to (\ref{MILocalaprox}). The integral is on the half sphere excepting an angle $\theta\sim \max(\epsilon/R,\epsilon \,m /R)$ from the AdS boundary. Therefore in the short distance limit $\epsilon/R\ll 1$ the difference with (\ref{MILocalaprox}) is a subleading term, and we get 
\begin{equation}\label{MILocalaprox1}
 I_{GFF}\sim k_{d+1} \, \frac{\pi^{d/2}}{\Gamma(d/2)}\,\frac{R^{d-1}}{\epsilon^{d-1}}\,,\hspace{1cm} \Delta \ll \frac{R}{\epsilon}\,.
\end{equation}    
In conclusion, we get a volume term with the same universal coefficient for any scaling dimension of the GFF, with the only provision that for large $\Delta$ the onset of the volume term is for large enough radius $R/\epsilon \gg \Delta \sim m$.

\subsection{First subleading term}

In the same way in which the leading term is a local contribution along the entangling surface in the bulk whose coefficient can be fixed thinking in a flat wall, the first subleading term in the short distance expansion comes from a boundary contribution in AdS, that grows with the area ${\cal A}$ of the region.  We will elaborate more on the general form of the expansion in the next section. Here we note that this same subleading term will be present in a configuration as the one computed by eq. \eqref{su}, when $A$ and $B$ have flat surfaces of area ${\cal A}$, close together at a distance $\epsilon\ll z_0$, and are extended in the bulk direction $z$ up to some arbitrary $z_0$. 

 We begin by recalling that the sum  in (\ref{su}) is over the masses $\kappa_n= z_0^{-1}\,j_{\alpha,n}$, where $j_{\alpha,n}$ are the zeroes of  $J_{\alpha}$. Except for low $n$, this can be fairly accurately approximated by a sum over positive integers as,
\begin{equation}\label{zeroesaprox}
   j_{\alpha,n} \sim n\pi + \frac{\pi}{4}(2\alpha-1)\,,\qquad \alpha > -1\,.
\end{equation}
We recall $\alpha=\Delta-d/2$.  
The contribution we are looking for is the one extensive in the area ${\cal A}$ for small $\epsilon$. In this geometric limit we can write generically
\be
 I_0(\kappa ,A,B)=f(\kappa  \,\epsilon)\,\frac{{\cal A}}{\epsilon^{d-2}}\,,
\ee
while for $d=2$ there is a logarithmic dependence on $\epsilon$.
 As the mass comes in a combination $\kappa  \,\epsilon$ we can consider an expansion in the perturbation $\delta=\frac{\epsilon}{z_0}\frac{\pi}{4}(2\alpha-1)\ll1$.  Crucially, $\delta$ does not depend on $n$.
 We find
\beb
I_\varphi(A\times I,B\times I) &=& \frac{{\cal A}}{\epsilon^{d-2}}\sum_{n=1}^{\infty}  f(\kappa_n\,\epsilon) \,,\\
\sum_{n=1}^{\infty}  f(\kappa_n\,\epsilon) &\sim& \sum_{n=1}^{\infty}  f(n \pi \,\epsilon/z_0) + \delta\,  \sum_{n=1}^{\infty} f'(n \pi\,\epsilon /z_0) + O(\delta^2)\,.
\eeb
We approximate both sums via the Euler-Maclaurin formula,
\begin{align}
     \sum_{n=1}^{\infty}  f(n \pi\,\epsilon/z_0) & \sim  \int_{1}^{\infty} dn \, f(n \pi\,\epsilon/z_0)-\frac{1}{2} f(\pi\,\epsilon/z_0)\sim  \frac{z_0}{\pi \epsilon} \int_{0}^{\infty} dx  f( x)-\frac{1}{2} f( 0 ) \,,\label{fte}\\
     \delta \,\sum_{n=1}^{\infty}  f'( n \pi \,\epsilon/z_0 ) & \sim \delta \int_{1}^{\infty} dn\;  f'( n \pi \,\epsilon/z_0 )+O(\delta^2)\sim -\frac{2\alpha-1}{4} f(0 ) + O(\delta^2)\;.
\end{align}
where we have used that $f(x)\to 0 $ as $x\to\infty$ and have replaced $f(\pi\,\epsilon/z_0)\rightarrow f(0)$. The first term in the right hand side of (\ref{fte}) gives the leading term analyzed previously, proportional to the bulk surface $z_0 \,{\cal A}$. This follows from \cite{Casini:2009sr},
\begin{equation}
 \frac{1}{\pi} \int_{0}^{\infty} dx  \,f( x) = k_{d+1}\,.
\end{equation}
The subleading terms instead depend on $f(0)=k_d$. They combine to give
\begin{equation}\label{MIsubleading}
   I_\varphi(A\times I,B\times I) - k_{d+1} \frac{{\cal A} \,z_0}{\epsilon^{d-1}} \sim  \begin{cases} -\frac{2\alpha+1}{12} \ln(z_0/\epsilon) +\dots & d=2 \\\\ -\frac{2\alpha+1}{4} k_d\,\frac{{\cal A}}{\epsilon^{d-2}} +\dots & d>2 \end{cases}
\end{equation}

In order to apply this result to the GFF we still have to identify what part of the contribution comes from the boundary at $z_0$. This corresponds to a Dirichlet boundary for a massless field, since in the $\epsilon\rightarrow 0$ the mass can be neglected for this boundary term inside the bulk (but not in the AdS boundary). Then this unwanted contribution is half the one of the case $\alpha=1/2$, that has two identical boundary conditions at the extremes of the interval $z\in (0,z_0)$. Subtracting this contribution, setting $\alpha=\Delta-d/2$, and replacing the area by the area of the sphere, we get
\be
I_{GFF}\sim k_{d+1} \, \frac{\pi^{d/2}}{\Gamma(d/2)}\,\frac{R^{d-1}}{\epsilon^{d-1}}\,+\,\begin{cases} -\frac{\Delta-1}{3} \ln(R/\epsilon) +\dots & d=2 \\\\ -(\Delta-d/2)\, k_d\,\frac{\pi^{d-1}}{\Gamma[d-1]}\frac{ R^{d-2}}{\epsilon^{d-2}} +\dots & d>2 \end{cases}\,,\hspace{1cm} \Delta \ll \frac{R}{\epsilon} 
\ee
For $d=2$ we have taken into account that the interval has two boundaries and we have to double the result of a single boundary.     

We can check the subleading term for $d=2$ matches the ones in Sec. \ref{SecConfBulk} for the conformal cases $\alpha=\pm1/2$ obtained from the conformal boundary anomalies. This result generalizes this terms for any mass or conformal dimension.

\subsection{General form of the short distance expansion}

The short distance expansion of the MI for an ``ordinary'' CFT follows the same pattern of the expansion of the EE where the cutoff is now replaced by the physical distance $\epsilon$ \cite{Casini:2015woa}. For a sphere this has the form
\be 
I= c_{d-2}\,\frac{R^{d-2}}{\epsilon^{d-2}}+c_{d-4}\,\frac{R^{d-4}}{\epsilon^{d-4}} +\cdots + \left\{\begin{array}{l} (-)^{d}\, 2 A\, \log(R/\epsilon)\hspace{.4cm} d\textrm{ even.} \\  (-)^{d}\, F \hspace{2.2cm} d\textrm{ odd.} \end{array}\right.
\label{CFTT}\ee
The last term is usually called the universal part, but for the MI all terms are universal in the sense that they are independent of the regularization. In particular, the first term has coefficient $c_{d-2}= \left(2\pi^{\frac{d-1}{2}}/\Gamma[\frac{d-1}{2}]\right)\, k_d$ proportional to the one of the mutual information between parallel planes. The logic of this expansion is that the terms divergent with $\epsilon$ are produced by local entanglement across the entangling surface and then are given by integrals of geometric quantities on the surface. To respect the symmetry between the MI between the inside and outside of the region these geometric terms are formed by even powers of the extrinsic curvature. Hence the powers involved in the expansion have the same parity as the area term \cite{Liu:2012eea,Grover:2011fa}. This also explains why there is a logarithmic term only in even dimensions. 

The conformal GFF differs notably from this expansion.  The arguments in the previous discussion can be extended to give the following expansion for the GFF
\be 
I_{GFF}= c_{d-1}\,\frac{R^{d-1}}{\epsilon^{d-1}}+c_{d-2}\,\frac{R^{d-2}}{\epsilon^{d-2}} +c_{d-3}\,\frac{R^{d-3}}{\epsilon^{d-3}} +\cdots + c_0\, \log(R/\epsilon) + \textrm{cons} \,,\hspace{.6cm} \Delta \ll \frac{R}{\epsilon}\,.\label{gfff}
\ee
Remarkably, the expansion starts at the volume term rather than the area term, and includes all integer powers of $R/\epsilon$. 
Again, divergent terms are produced locally by short distance correlations. However, in the present case the terms with the same parity of $d-1$ are coming from the bulk short distance entanglement while the ones with the parity of $d-2$ are local contributions associated to the boundary entangling surface.  

In particular, the leading term is the volume term  is (\ref{MILocalaprox1}) and the coefficient is independent of $\Delta$. Other terms with the same parity occur as subleading terms in the bulk short distance entropy expansion. For a massive scalar in a general $d+1$ dimensional curved space and a smooth shape of the entangling surface these are of the form\footnote{The coefficients for $a=b=0$ are known \cite{Casini:2014yca}.}
\be
\int \frac{dA}{(l(x))^{d-1-2\,(a+b+c)}} \, {\cal R}^a\,(K^2)^b\, (m^2)^c\,. \label{inte}
\ee
 These contributions are perturbative around the massless flat case. Therefore they are of the form (\ref{basically}) where powers of the distance $l(x)$ (the cutoff if we think in the entropy) are replaced by powers of the mass, the spacetime curvature ${\cal R}$, or the intrinsic curvatures $K$. 
When we plug this expression for our semi-spheres in AdS we have to take again $z\gg \epsilon, \epsilon m$. For smaller $z$ the approximation breaks down and the contribution is associated to the boundary. Integrating (\ref{inte}) along the bulk entangling surface gives us a term proportional to $(R/\epsilon)^{d-1-2(a+b+c)}$ in (\ref{gfff}). As a bonus we get that the coefficients $c_{d-3}, c_{d-5},\cdots$ can only depend on $\Delta$ through a polynomial in the mass square $m^2(\Delta)=\Delta (\Delta- d)$,
\be
c_{d-1-2 s}= a_{2s}\, (\Delta (\Delta- d))^{2 s} +a_{2s-2}\, (\Delta (\Delta- d))^{2 s -2} +\cdots \,. 
\ee
In particular, these coefficients will be the same for $\Delta=\frac{1}{2}(d\pm\sqrt{d^2+ 4 m^2})$ when these two solutions exist, $-d^2 /4\le m^2< -d^2/4+1$. We then expect the logarithmic coefficient $c_0$ for $d$ odd will be equal to the trace anomaly (\ref{odd}) only for the conformal bulk case $\Delta=(d\pm 1)/2$. For other scaling dimensions a polynomial dependence on the mass is expected.  

On the other hand, terms with the parity of $d-2$ appear as local contributions associated to the geometry of the boundary entangling surface. We have computed the first of such terms in the last section above. 
Because short distance in the boundary involves arbitrary long distances in the bulk, these contributions are non perturbative in the mass and the bulk curvature. They also depend on the boundary conditions. Then we expect the coefficients $c_{d-2}, c_{d-4},\cdots$ to be non polynomial functions of the mass. In particular the leading term is proportional to $\Delta-d/2=\pm\sqrt{d^2+ 4 m^2}$.    
 
As a conclusion, eq. (\ref{gfff}) violates the expected parity structure of a CFT MI (\ref{CFTT}). This is not due to any violation of the parity of the entropy between the inside and outside of the region but rather the consequence of the existence of divergent contributions that are not localized on the entangling region from the point of view of the $d$-dimensional theory. There is a double origin to the large correlations that give place to divergent terms. While the holographic description makes this structure quite transparent, the new large correlations residing in the bulk of the $d$-dimensional region are more difficult to grasp in terms of the GFF itself.   

The changes that occur for non spherical regions are then simple to track. For example, the volume term has the same structure, but now there is also a shape factor that takes into account the shape of the bulk entangling surface. As such it directly depends on the algebra choice. The area term remains unchanged however, and is still independent on the algebra choice.  

\section{Long distance mutual information \label{long}}
In this section we will consider the MI for two regions $A$ and $B$ in the long distance limit. When the regions $A, B$ are double cones the leading long distance term is universal for any CFT and applies as well to the case of a conformal GFF. We review this result below. The holographic description of the GFF for specific values of $\Delta$ gives a free conformal bulk. This leads to interesting holographic relations for the coefficients of the long distance MI for general CFT's in different dimensions. 
When the shape of $A$ and $B$ are not spheres, the coefficients of the expansion depend on further details of the theory. For the GFF, the long distance coefficients  depend on the chosen algebras for the regions. We analyze particular ``pinching'' limit for the shapes of the regions.  These are a test of global properties of the theories:  causality and interacting versus free UV limits.

\subsection{Two spheres}
The MI between two distant regions can be computed using an OPE expansion of the Renyi twist operators. The leading term comes from the lowest primary operator of dimension $\Delta$ in two replica copies \cite{Cardy:2013nua}. Hence the fall off power of the MI is $L^{-4 \Delta}$. There is a closed formula for the coefficient of this term in any CFT that depends on the algebras attached to these two regions through the action of the modular flow on the two point functions of the primary operators \cite{Casini:2021raa}. If the regions are spheres the modular flow in a CFT has a universal geometric expression which in radial coordinates is independent of the spacetime dimension \cite{Hislop:1981uh,Casini:2011kv}. The two point function only depends on the spin and $\Delta$. Therefore, the leading contribution of the MI of spheres is universally given as a function of the spin and $\Delta$ of the lowest dimensional primary, in a way independent of the space-time dimension and other details of the CFT \cite{Agon:2015ftl,Casini:2021raa}. In particular, these formulas must be valid for conformal GFF too. While these GFF do not have a stress tensor they are still CFT's, and the modular flow for spheres is still given by the usual one parameter group of conformal transformations leaving the sphere fixed. These conformal transformations are a symmetry of the theory even if they do not have a local expression in terms of the stress tensor \cite{dutsch2003generalized}.

When the lowest dimension primary is a scalar field the result is \cite{Agon:2015ftl}
\be 
I(A,B) \sim c\left(\Delta\right)\, \frac{R_A^{2\Delta}R_B^{2\Delta}}{L^{4\Delta}}\,,\qquad c\left(\Delta\right) = \frac{\sqrt{\pi}\,\Gamma[1+2\Delta]}{4\, \Gamma[3/2+2\Delta]}\,,
\label{mutualscalar}
\ee
where $R_A, R_B$ are the radius of the two spheres and $L$ the separating distance. 

\begin{figure}[t]
\centering
\includegraphics[width=.50\linewidth]{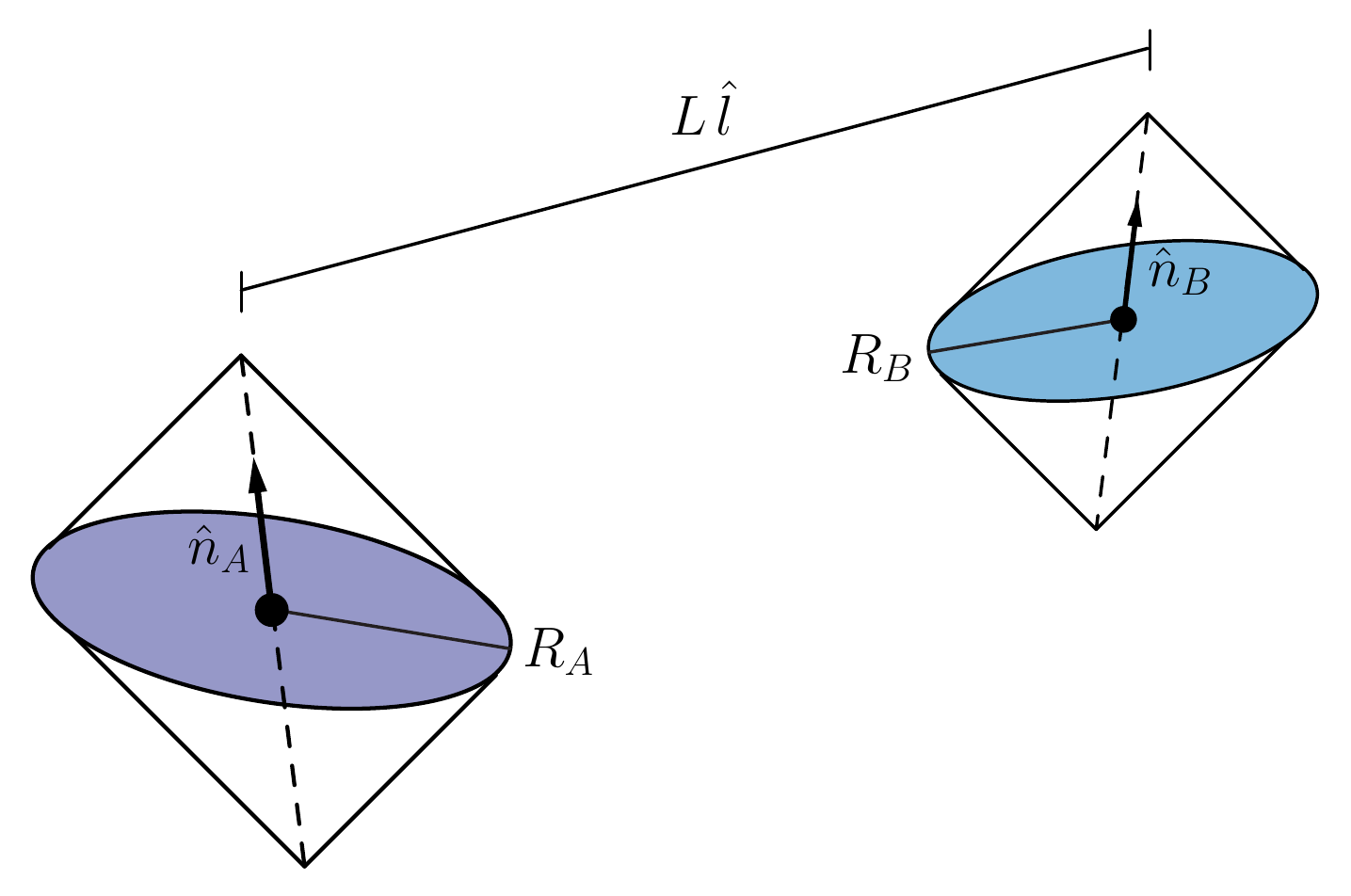}
\caption{\small Setup of two boosted spheres of radius $R_A,\, R_B$ and orientation $\hat{n}_A,\, \hat{n}_B$. Their separation is given by the vector $\vec{L}=L\,\hat{l}$ with $L \gg R_A, R_B$ .}
\label{boosted}
\end{figure}

One can also compute the long distance MI between spheres for fields of higher spin. The spin will introduce a dependence on the orientations of the double cones. We parametrize the geometry as follows. The space-time orientation of the double cones is given by the future directed time like unit vectors $\hat{n}_A,\hat{n}_B$ pointing in the direction of the vectors joining the tips of the double cones. See Fig.\ref{boosted}.  We write the spacial vector separating the sphere centers $L\, \hat{l}$.  A useful parameter describing this relative orientation is
\be 
\cosh(\beta) \equiv 2 (\hat{n}_A\cdot \hat{l}) (\hat{n}_B\cdot \hat{l})- (\hat{n}_A\cdot \hat{n}_B)\,.
\ee
Then, the leading term for the MI dominated by a Dirac spinor primary in $d$ dimensions is \cite{Casini:2021raa}, see also \cite{Chen:2017hbk}
\be 
I(A,B) \sim 2^{\lfloor\frac{d}{2}\rfloor+1}\, c(\Delta)\, \frac{R_A^{2\Delta}R_B^{2\Delta}}{L^{4\Delta}}\,\cosh(\beta)\,.
\label{mutualfermion}
\ee

Further Lorentz representation of the primary field in arbitrary dimensions  are described by a Young diagram of the symmetry of the tensor indices (for bosonic fields).  
For odd dimensions $d=2 q+3$, $q=0,1,\ldots$, the Young diagram giving the representation of the Lorentz group is determined by the lengths $m_1,\ldots,m_{q+1}$ of the rows, with $0\le m_1\le m_2\le \ldots m_{q+1}$.\footnote{If the number of rows in the Young diagram is less than $q+1$ we have to complete the sequence with zeros.} Defining the matrices 
\be
  A_j = 
\left\lbrace\frac{(m_s+s-1/2)^{2 p-1}}{(2 p-1)!}\right\rbrace_{p=1,\ldots,q}  
^{s=1,\ldots, \hat{j}
,\ldots , q+1 } \,,\hspace{.5cm} (\textrm{for}\, d=3,\, q=0,\, A_1=\{1\})\,,
\ee
where $\hat{j}$ means the index $j$ 
 is to be omitted,  we have for the leading term of the MI dominated by a primary with this representation:
\be\label{eq:oddI}
I(A, B)  \sim\, c(\Delta)\left(\frac{R_A R_B}{L^2}\right)^{2\Delta}  \left(\frac{\sum_{j=1}^{q+1} (-1)^{j+q+1}\,\sinh(2 \beta (m_j+j-1/2))\, \det(A_j) }{\sinh(\beta) (\cosh(2 \beta)-1)^q}\right)\,. 
\ee

For even $d=2 q+2$, $q=1,2,\ldots$ we have instead
\beb
A_j & = &  
\left\lbrace\frac{(m_s+s-1)^{2 p}}{(2 p)!}\right\rbrace_{p=0,\ldots,q-1}  
^{s=1,\ldots, \hat{j},\ldots , q+1 } \,,\\
I(A, B)  &\sim&\, c(\Delta)\left(\frac{R_A R_B}{L^2}\right)^{2\Delta}  \left( \frac{  \,\sum_{j=1}^{q+1} (-1)^{j+q+1}\,\cosh(2 \beta (m_j+j-1))\, \det(A_j) }{(\cosh(2 \beta)-1)^q}\right)\,. \label{iff}
\eeb
For even dimensions, if $m_1\ne 0$, we have two dual representations with the same tensor structure. If both components are present we have to multiply (\ref{iff}) by $2$. 

As mentioned, formulas (\ref{mutualscalar}), (\ref{mutualfermion}), (\ref{eq:oddI}), and (\ref{iff}) also give the leading long distance MI between spheres for conformal GFF of any conformal dimension and spin. These are defined as Gaussian fields with the two point function given by the unique conformal two point function of a primary field with the conformal dimension $\Delta$ and given Young diagram.  

The fact that the formula for the scalar contribution (\ref{mutualscalar})  depends  on $\Delta$ and not on the dimension $d$ is important to the consistency of the holographic description \cite{Agon:2015ftl}.  This is because the computation for GFF can be done independently both in the bulk and the boundary. When the regions $A$ and $B$ are far apart they can be considered  near boundary regions in AdS. Therefore, there is an agreement between the bulk and boundary fall-off of two point functions. On the other hand, the geometric action of the boundary modular flow for spheres coincide with the one of the bulk modular flow for semi-spheres attached to the boundary, essentially by symmetry reasons.  As mentioned above, the long distance contribution depends only on these elements, the two point function and the modular flow, and this gives the agreement of the boundary and bulk calculations for the GFF MI. 
 
Here we want to highlight interesting consistency relations for these general formulas for the mutual information that arise through GFF holographic descriptions in a like manner. Note that for fields with spin, given a specific Young diagram, the MI for boosted spheres does depend on the space-time dimension through the various terms in the spin structure. 
   
We will take a GFF with specific conformal dimension such as it is dual to a free conformally coupled field in AdS. Then we can eliminate the AdS metric by a Weyl re-scaling. This gives us an equation between the MI of the GFF in dimension $d$ and a free massless dual field in $d+1$ Minkowski space. We have to take into account that the regions in the bulk are now half-spheres instead of spheres and there is a conformal boundary condition at the boundary. This boundary condition  can only change the multiplicity of degrees of freedom because it does not change the modular flow.  

For example, there must be an equality between the MI of a massless free fermion field in $\mathbb{R}^{d+1}$, having $\Delta=d/2$, and a GFF spinorial field in $\mathbb{R}^{d}$ with the same dimension, because they are holographic dual to each other. The fermion contribution (\ref{mutualfermion}) has the dimension dependent factor $2^{\lfloor\frac{d}{2}\rfloor+1}$. This factor implies that for even $d$, the mutual information of the spinor in $\mathbb{R}^{d+1}$ matches the result for the GFF spinorial field in $\mathbb{R}^{d}$, whilst an extra factor of $2$ appears for odd $d$.
A naive counting of degrees of freedom reveals that the natural result is actually the mismatch for odd $d$: massless Dirac spinors in $\mathbb{R}^{d+1}$ have double the degrees of freedom of Dirac spinors in AdS$_{d+1}$, because the chiralities couple at the conformal boundary of AdS. Thus, it is actually the even $d$ result which needs and explanation for a missing factor of $2$ in the MI. This is directly linked to the dimensions of the $\gamma$ matrices and, in turn, in the way boundary fermions couple to bulk fermions via holography. For even $d$, Dirac AdS$_{d+1}$ couple to CFT$_{d}$ Weyl spinors \cite{Mueck:1999nr}, so in order to correctly compare Dirac GFF fermions in $\mathbb{R}^{d}$ to Dirac free fermions in $\mathbb{R}^{d+1}$ one should double the holographic result, providing the missing factor of $2$.

We can also consider other spin representations of the Lorentz group. Here the holographic identity will also involve a factor $2$ in the MI due to boundary condition. However, the identity is more interesting since it relates fields in different Young tableaux representations for boundary and bulk. In order to relate CFT's in the two different dimensions we need to find free conformal primary fields in $d+1$ dimensions. It is known that in addition to scalar and Dirac fermions there are free primaries only in even dimensions \cite{siegel1989all,minwalla1998restrictions}. This implies $d+1$ even and then odd $d$. The free primaries in $d+1$ dimensions have a Young tableau structure given by a rectangular diagram  with $(d+1)/2$ rows and $m$ columns. Hence there is a free primary for each $m$. It has dimension $\Delta=(d+ 2m-2)/2$.
 For example in $d+1=4$ these are 
\begin{figure}[H]
\centering
\includegraphics[width=10cm]{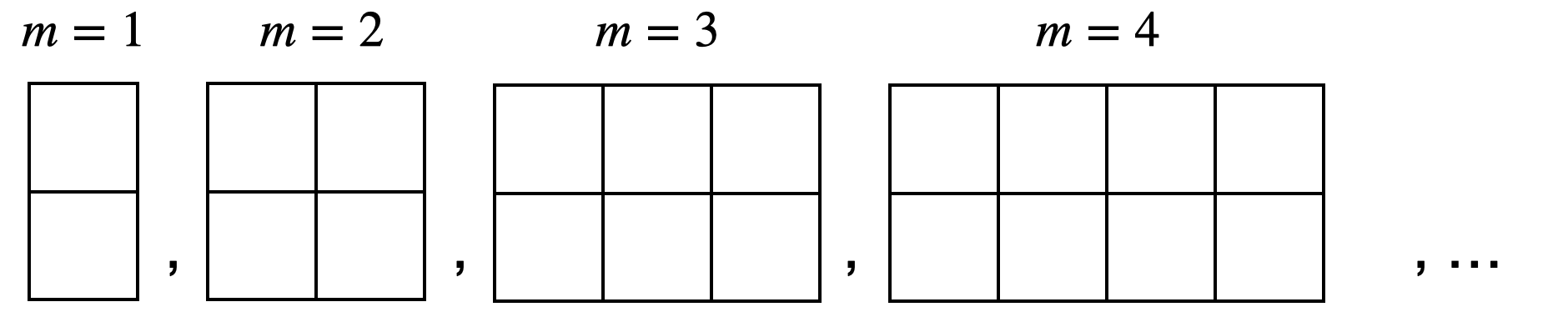}
\end{figure}
\noindent
The first corresponds to a Maxwell $F_{\mu\nu}$ field, the second to the curvature of a free graviton $R_{(\mu\nu)(\alpha\beta)}$, an so on. The label $m$ turns out to be the helicity of the particles. 
The MI in the long distance limit (\ref{iff}) is given by
\be 
I(A,B) \sim 2\, c(\Delta)\, \frac{R_A^{2\Delta}R_B^{2\Delta}}{L^{4\Delta}} \left[\frac{\cosh(2\beta(m+1))-\cosh(2\beta m)}{\cosh(2\beta)-1} \right] \label{id4}\,.
\ee
These fields are dual to fields in $d$ dimensions with the same $\Delta$ and Yang Tableaux that are obtained by taking out one of the rows. For example if $d=3$ we get    
\begin{figure}[H]
\centering
\includegraphics[width=10cm]{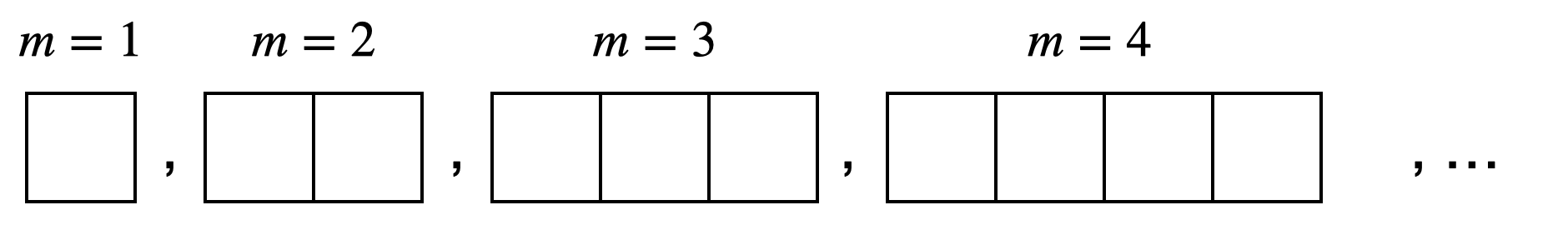}\,.
\end{figure}
\noindent 
For $m=1$ this is a current ($\Delta=2$), for $m=2$ the stress tensor ($\Delta=3$), and higher conserved currents for larger $m$.    
The MI in the long distance limit (\ref{eq:oddI}) writes in this case 
\be 
I(A,B) \sim c(\Delta)\, \frac{R_1^{2\Delta}R_2^{2\Delta}}{L^{4\Delta}} \left[\frac{\sinh(\beta(2m+1))}{\sinh(\beta)} \right] \label{id3}\,.
\ee

It is not difficult to show that (\ref{id3}) and (\ref{id4}) are in fact the same functions, except for the global factor $2$ appearing in (\ref{id4}). This difference appears because only half of the degrees of freedom of the free field in the larger dimension survive due to the boundary condition. 

It can be checked the same holds true if we consider any odd $d$. It is interesting to observe that, though there is nothing intrinsically holographic in  the general formulas for the leading term in the MI, these relations for different Young diagrams would have been difficult to spot without holography. However, the relation does not hold any more for even $d$, and this is because even if the $d+1$ dual free fields exist, they are not conformal primaries, and then we cannot arrive to relations between contributions for CFT's. Another commentary is that even if we have deduced that (\ref{id3}) and (\ref{id4}) should be proportional for the specific $\Delta$ of a free $d+1$ field, the relation between the MI contributions is still valid for any $\Delta$. However, in this case it is not obvious that it is expressing some form of holographic identity.

\subsection{Long distance MI under pinching }

When the regions are not spherical the modular flow for a CFT is not universal and geometrical.  In consequence details of the theory that go beyond the lowest primary spin and dimension are revealed in the long distance MI. In particular, for the GFF, the choice of algebra becomes relevant.  

\begin{figure}[t]
\centering
\includegraphics[width=.4\linewidth]{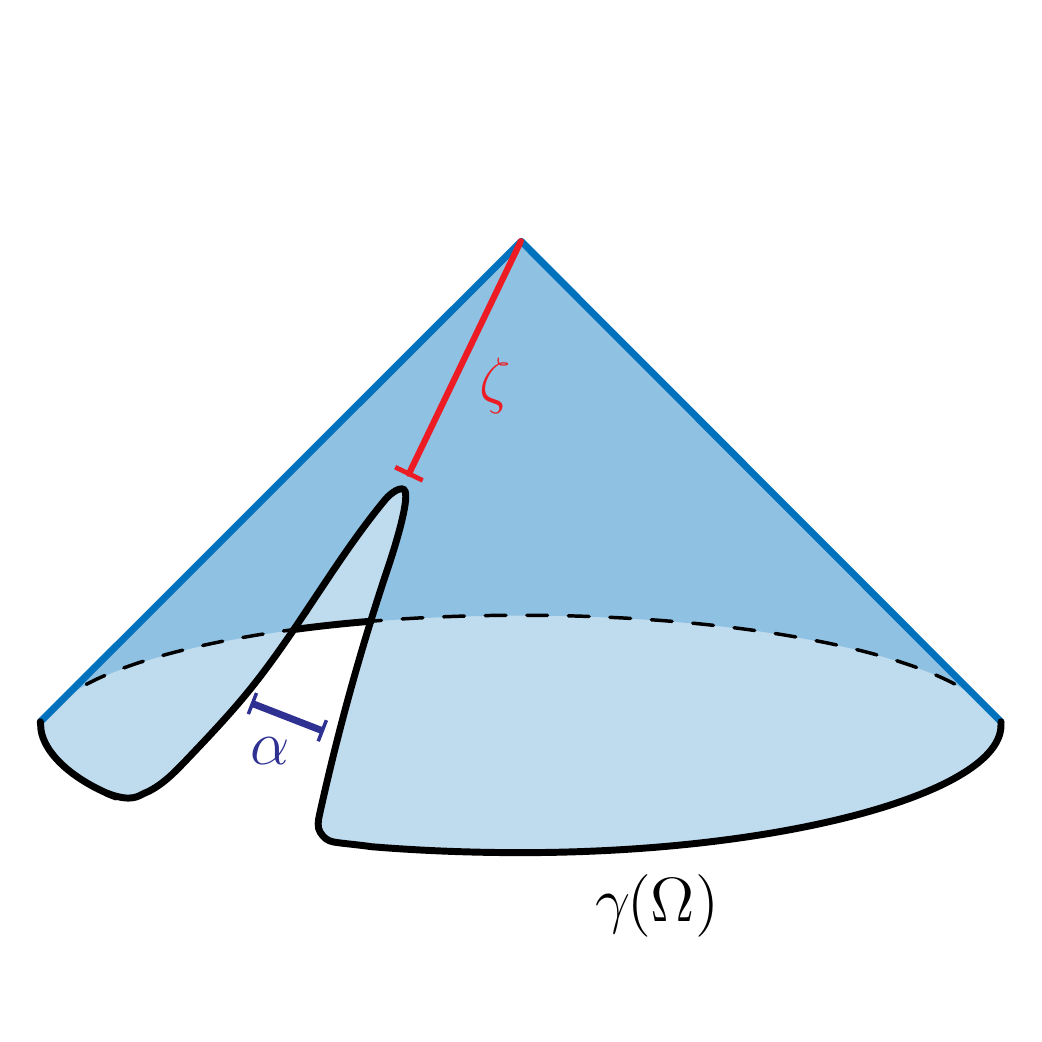}
\caption{ \small Curve $\gamma(\Omega)$ defining the pinching over the  future horizon of the double cone. The curve is described by two parameters: ${\alpha}$ which is related to the thickness of the region removed from the null cone and  ${\zeta}$ with its height. }
\label{pinchfig}
\end{figure}

We start from a similar set-up as in Fig. \ref{boosted} but with two spheres in the same Cauchy slice (i.e. unboosted) of radii $R_A,R_B$. We will leave one of the spheres untouched, say $B$. Instead of a sphere $A$ we take a causal region with boundary described by a curve $\gamma(\Omega)$ on the future horizon of the double cone $A$. $\Omega$ are the angle variables describing the directions on the cone, and $\gamma(\Omega)$ is the radial (or temporal) coordinate of the boundary. For a sphere of radius $R_A$, the curve $\gamma(\Omega)$ is constant $\gamma=R_A$. We choose $\gamma(\Omega)$ to be a curve determined by two positive parameters $\{\zeta,\alpha\}$ that essentially removes a piece of the horizon of the double cone along some of the null generators as shown in Fig. \ref{pinchfig}. We define $\zeta$ as the shortest distance between the apex of the cone and $\gamma(\Omega)$. The other parameter  $\alpha$ represents the width of the region where $\gamma(\Omega)$ differs appreciably from $R_A$. The limit $\alpha\to 0$ corresponds to a very narrow subtraction, whilst $\zeta\to 0$ corresponds to removing up to the tip of the cone. The particular way in which this curve is parametrized is not essential for our purposes, but these two parameters play an important role in diagnosing important properties of the algebras assigned to the regions. We refer to this geometric deformation as ``pinching'' the original double cone.

The limits of interest are $\alpha,\zeta\to0$ but the order of the limits is important. The case where the limit $\zeta\rightarrow 0$ is taken first was introduced in \cite{Casini:2021raa}. In this case a narrow strip is removed from the cone all the way to the apex, resulting in a system defined by a null surface, i.e. the causal development of the pinched surface has zero spacetime volume. In this limit, the mutual information will drop to zero unless the theory contains ordinary free fields in the algebra, satisfying a linear equation of motion (as opposed to other GFF). The reason is that smearing fields only on a null surface is not enough to produce an operator in the Hilbert space, unless the operator is free, and its scaling dimension saturates the unitarity bound  \cite{Schlieder:1972qr,Wall:2011hj,Bousso:2014uxa}. So this pinching limit eliminates the algebra and leads to zero mutual information in the non free case.
That is, we expect
\begin{equation}\label{diag-free}     
    \lim_{\alpha\rightarrow 0} \,\lim_{\zeta\rightarrow 0} \, 
    I(\gamma(\Omega),B)\begin{cases} =0 & \text{ interacting CFT}\\ > 0 & \text{ CFT contains a (decoupled) ordinary free field}\end{cases}
\end{equation}
Notice that this defines the conformal GFF as interacting since the correlator for $\Delta>(d-2)/2$ is too singular to allow operators localized  on the null surfaces.\footnote{We emphasize that in the comparison with large $N$ holographic models, the mutual information turning zero in our computation does not mean that $I\sim O(N^0)$. This is always true in the present set-up of long distance between the two regions, but rather a strict limit $I=0$ between the systems.} In this case the conformal symmetry fixes the leading contribution of the mutual information to vanish with a power law given by a pinching exponent $\lambda$ as
\be
 I(\gamma(\Omega),B)\sim \left(\frac{\zeta}{\alpha}\right)^{\lambda}\,,\quad \lambda >0\,,\,\,\, \alpha,\zeta\to0 , \,\,\, \alpha\gg\zeta\,.
\ee
However, notice that this way of discriminating free from interacting theories cannot be used for $d=2$ where there is always a non trivial algebra in null intervals. 

Taking the limit $\alpha\rightarrow 0$ first corresponds to removing a single null segment from the null cone. This might be seen as containing no information (we are essentially removing a region of measure zero) on the smeared fields algebra, and thus the result should be exactly the same as for the sphere. However, The limit $\alpha\rightarrow 0$ of the causal region determined by $\gamma(\Omega)$ is not the causal region determined by the sphere, i.e. the double cone, but a smaller space-time region. This limit is then not a causal region, see Fig. \ref{elipsefig} below. Its causal completion coincides with the double cone however. Theories (or algebra assignations) satisfying primitive causality\footnote{Primitive causality assets that the algebra of a time like cylinder is equal to the one of its causal completion.} will in principle not notice the difference, because in that case the limit of the algebras should converge to double cone algebra. Then, we expect   
\begin{equation}\label{diag-causal}
    \lim_{\zeta\rightarrow 0} \, \lim_{\alpha\rightarrow 0} \,
    I(\gamma(\Omega),B)\begin{cases} =I(A,B) & \text{ causal theory}\\ < I(A,B)  & \text{ non causal theory }\end{cases}
\end{equation}
By ``causal'' in this context we mean that given the initial data of a subsystem one should be able to recover the complete causal diamond, even if we remove any number of zero-measure regions.\footnote{A related property has been called strong additivity \cite{Kawahigashi:1999jz}.} As we will see, the GFF behaves as causal or non causal in this sense depending on the algebra choice. 

Once again, we will study the GFF system via holography. To have a well defined problem we must choose the net of algebras associated to each region in the CFT. As explained before, this maps via holography to choosing a bulk wedge dual to the CFT subsystem. For a single region, the smallest consistent bulk region is the causal wedge and the biggest one is the complementary region's causal wedge. An intermediate choice that can be made consistent for all regions at the same time corresponds to the entanglement wedge coming from the HRT prescription. In this section, we will explore the behavior of \eqref{diag-free} and \eqref{diag-causal} both for the causal wedge and for the entanglement wedge. We will see that both MI computations allow to clearly distinguish the algebras. Mathematically, while the causal wedge relies on a causal geometric construction, the entanglement wedge arises as a solution to a differential equation. This amounts to the former being more sensitive to some deformations (e.g. removing the tip of the cone or not) than the latter which is more stable upon continuous deformations.

For quantitative analysis we consider the simplest case of a scalar free field in the AdS$_4$ with 
$\Delta=1$. In this case we are in the conformal bulk case, and we can map the problem from AdS$_4$ to $\mathbb{R}^4$ with conformal Neumann boundary conditions at $z=0$. This is performed by means of a Weyl transformation in a similar fashion as in previous sections, leaving the MI invariant. 

There is another reason for choosing the conformal case and take the region $\gamma$ with boundary on the null cone. In general, the coefficient of the long distance MI for arbitrary regions would be hard to compute, even for free fields. However, closed expressions are known for free primary fields when the region has boundary in a null cone \cite{Casini:2021raa}. In the present case it turns out that both the casual wedge and the entanglement wedge corresponding to $\gamma$ have boundaries that lie on the bulk null cone that corresponds to the boundary one. The reason for the simplification of the coefficient in this case is that the modular flow of regions in the null cone is local on the different null generators. In the free case this gives an expression for the coefficient that is a local integral on the region boundary \cite{Casini:2021raa}. The existence of a conformal boundary condition does not change these features. For a scalar this is proportional to the area of the boundary   
\beb
&{}&I(A,B)= C(1)\,\frac{C_A\, C_B}{L^{4}}\,,  \quad C_{A,B} = \frac{1}{2\pi}\,\int_{\Sigma_{A,B}} d\sigma_{A,B}\,,\label{ddd}
\eeb
where $C(1)$=4/15, and  $\Sigma$ is the boundary of the region in the bulk corresponding to the curve $\gamma$. For $C_B$ we consider a perfect hemisphere and therefore obtain $C_B = R^2_B$. 
\subsubsection{Pinching the entanglement wedge in AdS$_4$/CFT$_3$ }\label{pinch1}
For calculation convenience we will introduce first a region with boundary  in the null plane and then we will map it to the null cone. Consider coordinates $\tilde{x}^\pm=\tilde{x}^0\pm \tilde{x}^1$ and $\tilde{y}=\tilde{x}^2$ in $d=3$. The curve $\gamma$ on the null plane $\tilde{x}^-=0$ is defined by
\be 
\tilde{x}^-=0\,,\,\, \tilde{x}^+ = {\tilde{\gamma}} (\tilde{y})\,,\,\,  \,\, \tilde{y} \in \mathbb{R}  \,. \label{dV}
\ee
We choose a simple Lorentzian curve to study the pinching limit 
\be 
{\tilde{\gamma}} (\tilde{y}) =R_A \frac{(R_A-{\zeta})}{{\zeta}}\,\left( \frac{{\alpha}^2}{{\alpha}^2 +\tilde{y}^2} 	\right)\,.
\label{gammaL}
\ee
The parameter ${\alpha}$ is related to the thickness or the region removed from the null cone and  ${\zeta}$ to the height.  At ${\zeta}\to 0$ the pinching is complete and ${\alpha}\to 0$ is the thin limit.

Now, we can compute the corresponding 2-dimensional HRT surface $\tilde{\Sigma}$ in the bulk using light-cone coordinates $ds^2=\tilde{z}^{-2} (- d\tilde{x}^+ d\tilde{x}^- +d\tilde{y}^2+d\tilde{z}^2)$.
 The solution lies in the $\tilde{x}^-=0$ null plane in the bulk. The surface obeys a differential equation that can be solved by Fourier transformation \cite{Casini:2018kzx}. We get 
\be
\tilde{x}^-(\tilde{y},\tilde{z})=0\,,\quad \tilde{x}^+(\tilde{y},\tilde{z}) =\int^\infty_{-\infty} dk\,\, a_k \,e^{ik\tilde{y}}\,e^{|k|\,\tilde{z}}\,(1-|k| \tilde{z})\,,
\label{reg}
\ee
where the weights $a_k$ are given by the conditions imposed over the boundary surface $\tilde{x}^+(\tilde{y},\tilde{z}=0) = {\tilde{\gamma}} (\tilde{y})$ as
\be 
a_k = \int^\infty_{-\infty} \frac{d\tilde{y}}{2\pi}\, e^{-ik\tilde{y}}\, {\tilde{\gamma}} (\tilde{y})={\alpha} \,R_A \frac{(R_A-{\zeta}) }{2 \,{\zeta}}e^{-{\alpha}  |k|}\,. \label{ak}
\ee
The surface in the bulk parametrized by $y$ and $z$ can be obtained replacing (\ref{ak}) in (\ref{reg}) 
\be
 \tilde{x}^+(\tilde{y},\tilde{z}) = \, R_A \frac{(R_A-{\zeta})}{{\zeta}}\left[\frac{\tilde{y}^2{\alpha}^2+(\tilde{z}+{\alpha})^2(2\tilde{z}+{\alpha}){\alpha}}{(\tilde{y}^2+(\tilde{z}+{\alpha})^2)^2}\right]\,.
 \label{pinchplane}
\ee
The conformal map between the null plane to the null cone  extends to an isometry of AdS from coordinates $\tilde{x}^\mu\equiv (\tilde{t}, \tilde{x},\tilde{y},\tilde{z})$ to the ones $x^\mu\equiv (t,x,y,z)$. This is explicitly given by
\be 
x^\mu =\frac{2(\tilde{x}^\mu + (\tilde{x}\cdot \tilde{x}) C^\mu) }{1+2\, (\tilde{x}\cdot C) + (\tilde{x}\cdot \tilde{x})(C\cdot C)}-D^\mu\,,
\label{iso}
\ee
where the dot $\cdot$ is the usual Minkowski scalar product and the parameters are 
\be 
C^\mu = (0,\,1/R_A,\,0,\,0)\,,\quad D^\mu = (R_A,\,R_A\,,0,\,0)\,.
\ee
Taking the limit to the boundary as $z\to 0$ or $\tilde{z} \to 0$ requires us to substract a  global factor,
\beb
&{}& ds^2 = -dt^2 + dx^2 + dy^2 =\Omega_A^2(\tilde{t},\tilde{x},\tilde{y}) (-d\tilde{t}^2 + d\tilde{x}^2 + d\tilde{y}^2 )\,,\label{map}\\ 
&{}& \Omega_A (\tilde{t},\tilde{x},\tilde{y}) = \frac{2 \, R_A^2 }{-\tilde{t}^2 + (\tilde{x}+R_A)^2 +\tilde{y}^2}\,.
\eeb
This implies that the transformation (\ref{iso}) is a conformal transformation over Minkowski space in the boundary theory. It maps the curve (\ref{dV}) on the null plane to the one on the null cone. Then, we can compute the minimal surface $\Sigma$ bounding the entanglement wedge for $\gamma$ on the null cone by applying the corresponding isometry (\ref{iso}) to the the bulk surface $\tilde{\Sigma}$ determined by 
\be 
\tilde{x}^-=0\,,\,\, \tilde{x}^+ = \tilde{x}^+(\tilde{y},\tilde{z}) \,,\,\,   \tilde{y} \in \mathbb{R}\,,\,\, \tilde{z} \in \mathbb{R}^+_0  \,. 
\ee
In radial coordinates defined by $r^\pm= \sqrt{x^2 + y^2 + z^2 } \pm t\,$, we  obtain the parametrization for the HRT surface to be 
\be 
 \begin{cases}
 	   &r^+(\tilde{y},\tilde{z})=0\\ 
       & r^-(\tilde{y},\tilde{z}) = \Upsilon_A (\tilde{y},\tilde{z}) \left(R_A +\frac{\tilde{y}^2+\tilde{z}^2}{R_A}\right) \\
       & y (\tilde{y},\tilde{z}) = \Upsilon_A (\tilde{y},\tilde{z})\tilde y \\
       & z (\tilde{y},\tilde{z}) = \Upsilon_A(\tilde{y},\tilde{z})\tilde z\\
     \end{cases}
\ee
where $\Upsilon_A(\tilde{y},\tilde{z})$ is given by 
\be 
\Upsilon_A(\tilde{y},\tilde{z})=\frac{2R_A^2}{R_A^2+R_A\, \tilde{x}^+(\tilde{y},\tilde{z})+\tilde{y}^2+\tilde{z}^2}\,.
\label{OA}
\ee
Note that $r^+(\tilde{y},\tilde{z})=0$ is expected as the surface is on the past null cone. Additionally, all the curves pass through the point $x^\mu=(-R_A,R_A,0,0)$ which corresponds to $z\to \infty$. A plot of these surfaces is shown on the left panel of Fig.\ref{wedges}.

The coefficient $C_A$ for the mutual information can be computed as in (\ref{ddd}). This yields  the final result for the MI as
\be
I_{GFF}^{EW}(\gamma,B)  =  \left(\frac{1}{2\pi\, R_A^2} \int_{\Sigma} d\sigma_A\right)\,I_0(A,B) = \left(\frac{1}{2\pi R_A^2} \int_{\tilde{\Sigma}} d\tilde{y} d\tilde{z} \,\Upsilon_A^{2}(\tilde{y},\tilde{z}) \right)\,I_0(A,B)\,\,,\label{MIcausalw}
\ee
where $I_0(A,B)= {4\,R_A^2\, R_B^2}/{15 L^4}$ is the mutual information  between the full spheres. We now compute the MI using this formula for the limits of interest. For small $\zeta$ we get 
\begin{align}\label{compare-ent}
   I_{GFF}^{EW}(\gamma,B)
     &\sim c\,\,I_0(A,B) \,\left( \frac{\zeta}{\alpha}\right)^{\frac13 }  \,\to 0\,,  \qquad\qquad \alpha,\zeta\to0 , \quad \alpha\gg\zeta\,.
\end{align}
where $c=0.3675...$ is a constant that can be computed numerically. The limit of small $\alpha$ is instead 
\begin{align}
    I_{GFF}^{EW}(\gamma,B)\sim
   I_0(A,B)  \, \left( 1-\frac{\alpha}{\zeta}\right) \,\to I_0(A,B)\,,  \qquad\qquad \alpha,\zeta\to0 , \quad \alpha\ll\zeta \,.
\end{align}

According to  \eqref{diag-free} and \eqref{diag-causal} these results indicate that the theory is interacting and that the choice of algebras is causal. This is reassuring since the mutual information for the GFF with the entanglement wedge coincides with the one of a large N limit of strongly interacting theory with stress tensor, respecting causality. 

\subsubsection{Pinching the causal wedge in AdS$_4$/CFT$_3$ }\label{pinchcausal}

For the analysis of the causal wedge we proceed as follows. The causal wedge is the bulk shadow of the boundary region. It is not difficult to realize that for fixed $\zeta\neq0$ and small $\alpha$ the causal region in the boundary does not change drastically between $\alpha\ll 1 $ and the limit $\alpha=0$. Then the same happens for the causal wedge in the bulk. 
Thus, for this case we can approximate the regime of interest by taking $\alpha=0$ and studying the $\zeta\to 0$ limit. 

\begin{figure}[t]
\centering
\includegraphics[width=.4\linewidth]{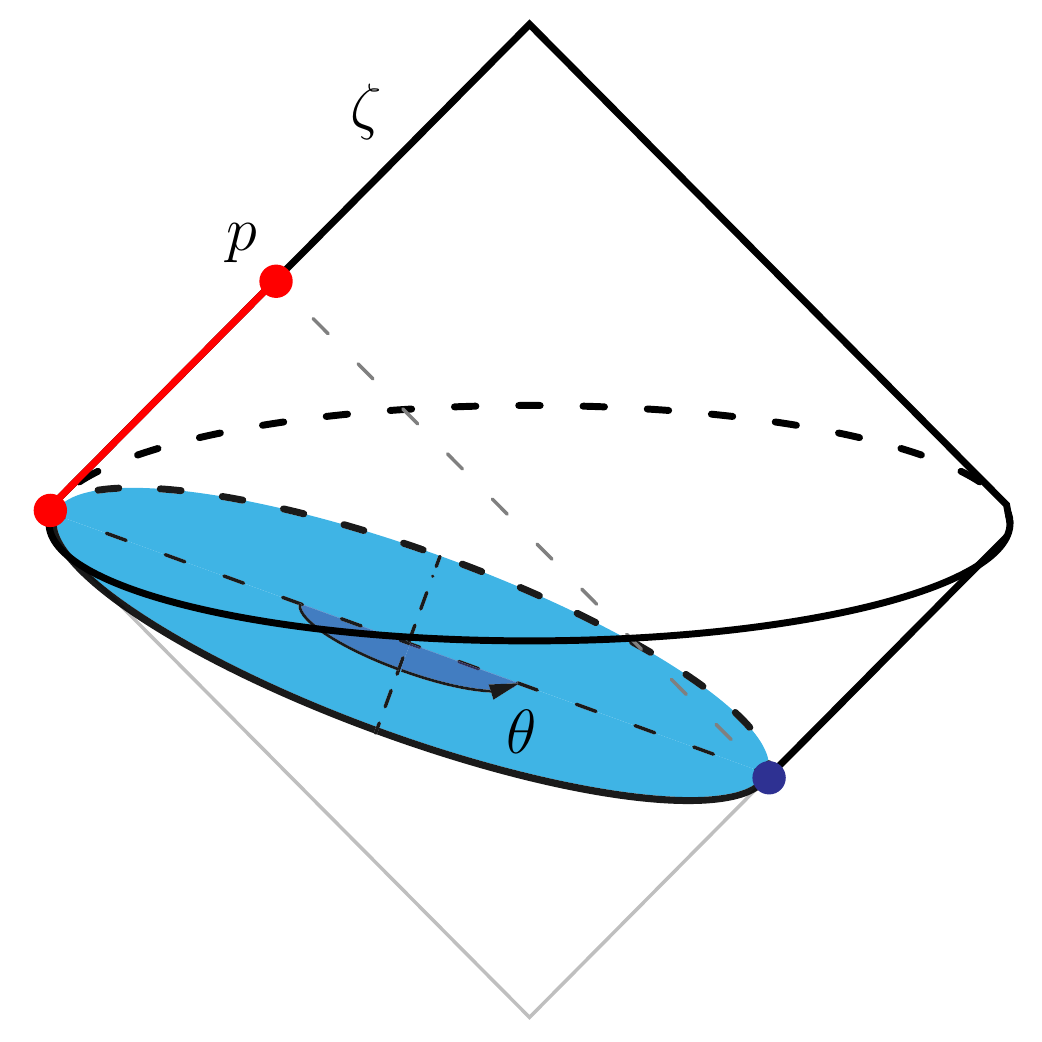}
\caption{\small Ellipse at the boundary of the causal set obtained by subtracting from the double cone on the CFT$_3$ the past of a single point $p$ at a distance $\zeta$ from the tip.}
\label{elipsefig}
\end{figure}

\begin{figure}[t]
\centering
\includegraphics[width=.45\linewidth]{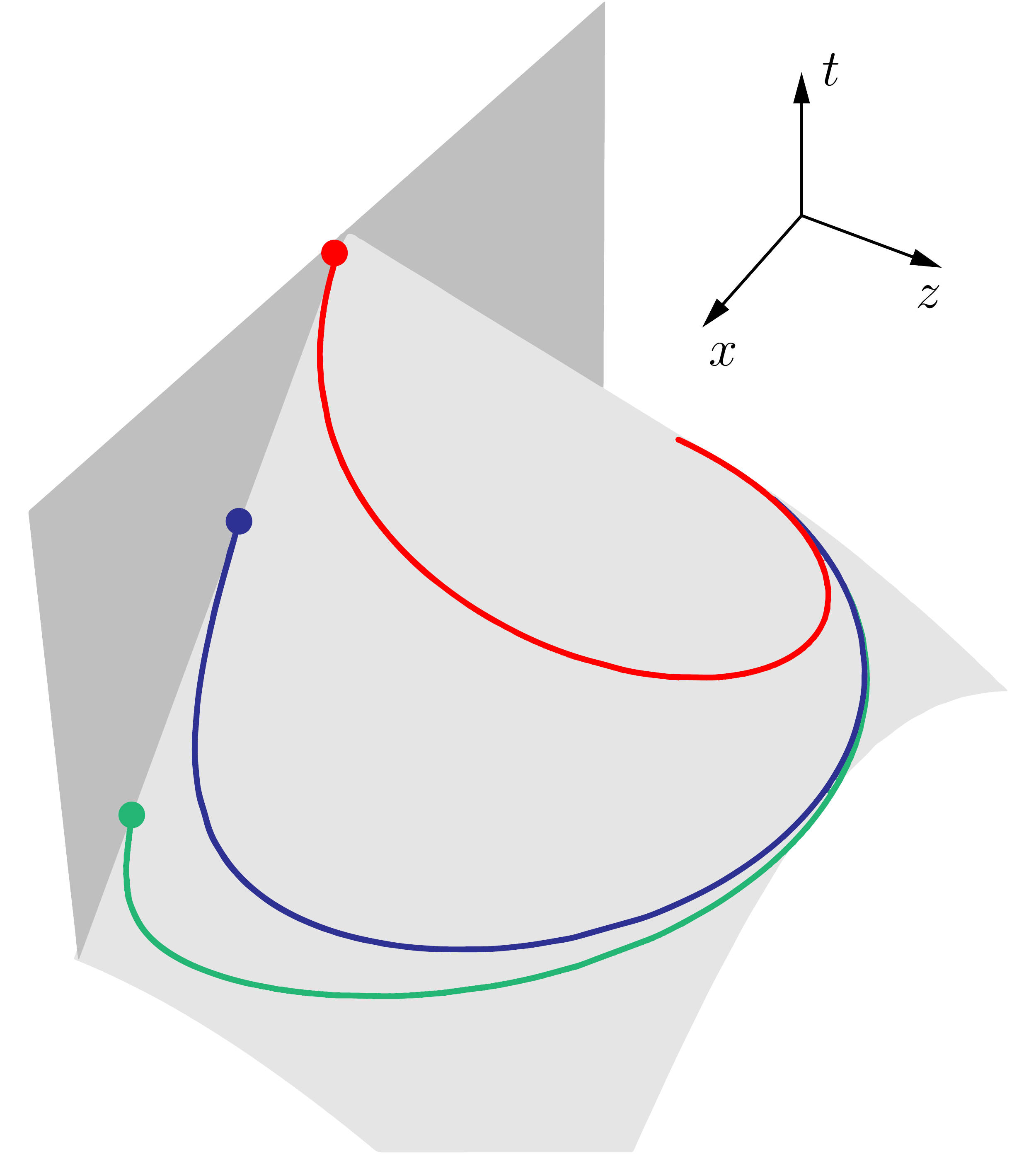}
\includegraphics[width=.45\linewidth]{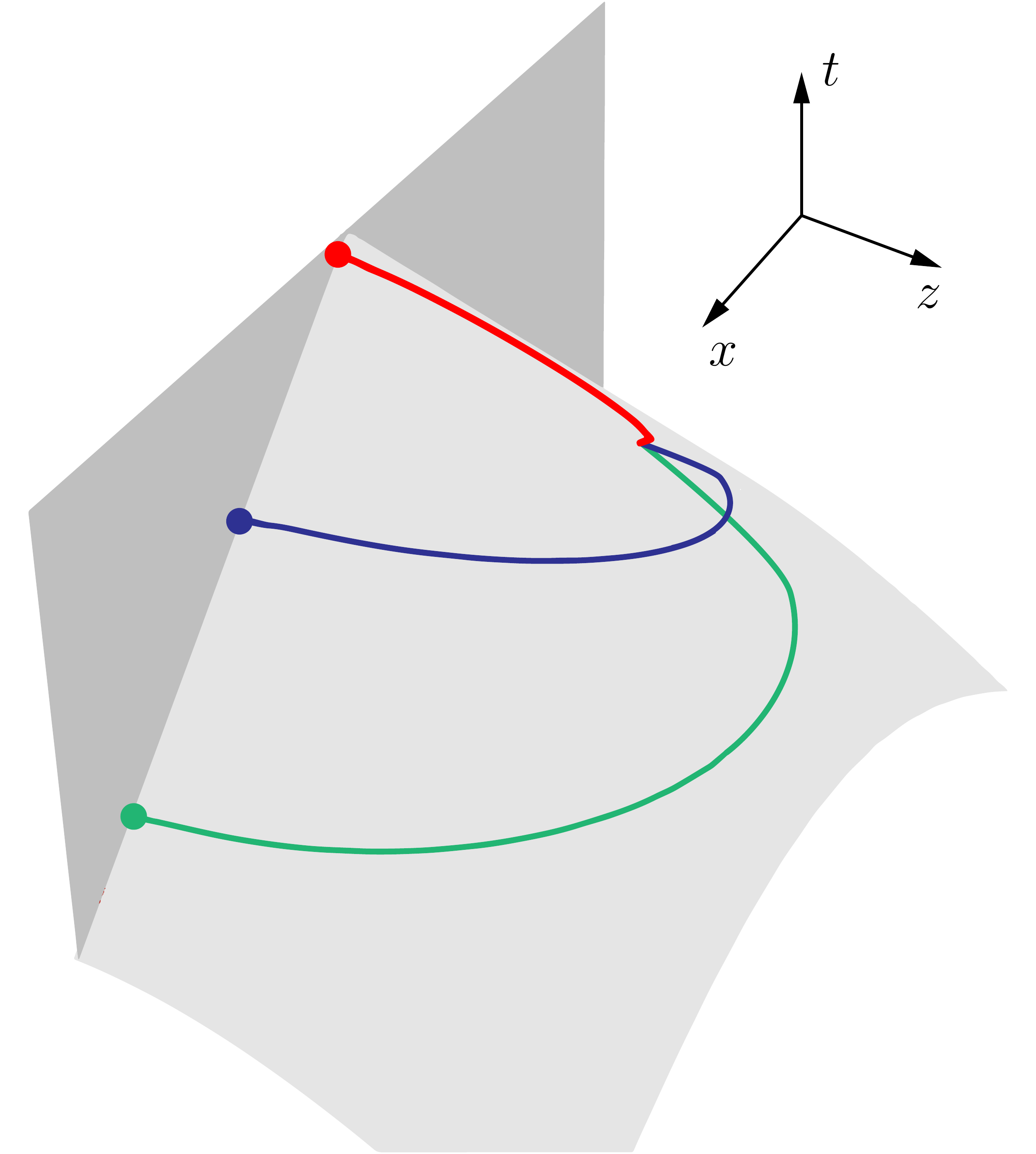}
\caption{\small Entanglement wedge for $\alpha=0.01$ (left)  and causal  wedge for $\alpha=0$ (right) on the $y=0$ plane. The  values of $\zeta$ for the pinching curves are given by  $0.03$  (red), $0.4$ (blue) and $0.8$ (green).}
\label{wedges}
\end{figure}

The $\alpha=0$ limit amounts essentially to remove a single null segment  from the future cone up to a distance $\zeta$ from the tip and consider the resulting bulk causal shadow of the region. This will differ in spacetime volume from the double cone even if the difference on the null Cauchy surface is of measure zero. The resulting causal set is obtained by subtracting from the double cone the past $J^-(p)$ of a single point $p$ at a distance $\zeta$ from the tip. The intersection between $J^-(p)$ and the past horizon of the double cone is given by the ellipse
\be 
 \begin{cases}
       & t(\theta)=-\left(R_A+\frac{\zeta }{2}\right)+\frac{\zeta}{2}  
   \cos (\theta )\,,\\
 	  & x(\theta)=\frac{\zeta }{2}+\left(R_A-\frac{\zeta }{2}\right)
   \cos (\theta )\,\\ 
      &  y(\theta)=\sqrt{R_A(R_A-\zeta) } \sin (\theta
   )\,.\\
     \end{cases}\qquad\theta\in(-\pi,\pi)
\ee
where $\theta$ is the angle that parameterizes the ellipse as in Fig.\ref{elipsefig} .
The origin of the time coordinate was put on the cone apex. Now, the causal wedge boundary is defined as the intersection between light-rays travelling into the bulk from the apex and from this ellipse. This problem can be solved for finite $\zeta$, giving a two dimensional surface where $\{t,x,z\}$ is parametrized as a function of $\theta$ and  a new angular variable $\phi$. The full expression is not very illuminating. The leading order for $\zeta\sim 0$ is
\be 
 \begin{cases}
 	   &r^+(\theta ,\phi)=0\\ 
       & r^-(\theta ,\phi) =  2 R_A \sin (\phi)  +2\zeta \left[1-\sin (\phi)\left( 1- \frac{\cos (\theta)}{2}\right) \right]
    \,, \\
       & y (\theta ,\phi) = R_A \sin{\theta}\sin{\phi} \\
       & z(\theta ,\phi)=\sqrt{\frac{2\,R_A\,\zeta\, \sin^2 (\theta)\,\sin (\phi)\, (1
   -\sin (\phi))}{1+\cos (\theta )}}\,.\\
     \end{cases}\qquad\theta\in(-\pi,\pi)\,,\,\,\,\phi\in(0,\pi/2)\,.
\ee
A plot of these curves is shown on the right panel of Fig.\ref{wedges} for some values of $\zeta$. This gives the holographic surface $\Sigma(\Omega)$. 

The computation of $C_A$, eq. (\ref{ddd}) yields
\begin{equation}\label{compare-caus}
 I_{GFF}^{CW}(\gamma,B) \sim  \frac{I_0(A,B)}{2}\,  \left(\frac{\zeta}{R_A} \right)^\frac{1}{2}\,, \quad \quad \zeta\ll 1\,.
\end{equation}
We conclude that the MI of the causal wedge $I_{GFF}^{CW}(\gamma,B)$ vanish as (\ref{compare-caus}) for $\zeta$ small, disregarding the value of $\alpha$. In the limit of $\alpha\rightarrow 0$ with $\zeta$ small we have a finite limit (\ref{compare-caus}) that is smaller than $I_0(A,B)$. This limit vanish if we further take $\zeta\rightarrow 0$. Therefore, the causal wedge describes a CFT algebra that has no free field sectors, and it is not causal in the sense defined above. The first is expected since for $\Delta\ge (d-2)/2$ no algebra can be localized in a null surface. The second statement is also reassuring in the sense that, the GFF not having a $T_{\mu\nu}$, its local algebra associated to a subsystem in a Cauchy slice may not be able to rebuild the causal diamond. This computation checks that this is the case in terms of the mutual information.

\subsubsection{Comparing the entanglement and causal wedges under pinching}

We now summarize by comparing the $I_{GFF}^{EW}(\gamma,B)$ and $I_{GFF}^{CW}(\gamma,B)$ results. Lets focus first on the $\alpha,\zeta\to0$, $\alpha\gg\zeta$ regime, \eqref{compare-ent} and \eqref{compare-caus}. The MI vanish in both of these computations. Geometrically this is because the bulk surface collapses to the AdS null boundary in the limit. This is consistent with the fact that the theory does not have any free field sectors. An extra constraint arises in the comparison, since it is known that the entanglement wedge algebra always contains the causal wedge one and this must be reflected in our computation by monotonicity of MI. This is reflected as an ordering in the pinching exponents 
\be
I_{GFF}^{CW}(\gamma,{\cal R}_B)\sim\zeta^{\frac 12} < \zeta^{\frac 13} \sim I_{GFF}^{EW}(\gamma,B)\,,\qquad\qquad \zeta\ll\alpha\ll R_A\,.
\ee
In the opposite regime $\alpha,\zeta\to0$, $\alpha\ll\zeta$, both quantities behave very differently,
\be 
I_{GFF}^{CW}(\gamma,B)\sim\zeta^{\frac 12}\,, \qquad\qquad I_{GFF}^{EW}(\gamma,B)\sim I_0(A,B)\,,\qquad\qquad \alpha\ll\zeta\ll1\,.
\ee
 This checks that the causal wedge alone does not contain the necessary operator content to reproduce the full double cone in the pinching limit whilst the entanglement wedge does. As mentioned this is a necessary feature of the entanglement wedge to match the MI of a complete theory with stress tensor. 
\noindent

\section{Final remarks}
GFF are quantum field theories with unusual properties. In this paper we have studied the manifestation of these unusual features in the MI. The most salient feature is a volume term in the MI instead of the usual area term. This does not happen for other theories associated to dimensional reduction such as Kaluza Klein models. In these models, the MI either has an area term if the interior space is of finite volume, or there is no split property and the MI diverges if the interior space has infinite volume. 
   
It is clear that the GFF show a form of bilocality, in the sense that large correlations between complementary regions are obtained near the boundary but also in the bulk of the region. These later are due to certain non local linear combination of fields in the causal region which holographicaly represent bulk fields living near the bulk entangling surface. This same combinations of fields must have also large correlations for other theories. The difference is that for ordinary theories, instead of being fresh new correlations, these field combinations in space-time can be written in a common Cauchy surface, and the large correlations should be in fact already counted by the area terms in the entropy. This points to an heuristic understanding of the origin of the large entropy of GFF as due to the existence of too many independent operators distributed for different times. In fact, for nearly complementary regions with a finite time span the GFF has only an area term in the MI, as can be easily seen holographically. For an ordinary QFT these operators at different times are assimilated to the same operators on the Cauchy surface by using the equations of motion.  

Holographic models avoid having a volume term by the existence of a phase transition well before the short distance limit of the GFF MI is achieved.      
 The large flexibility of algebra choices of the GFF is important for their holographic role since it allows to fake causality by the entanglement wedge algebra choice.  

As a consequences of the volume term, the usual irreversibility inequalities for the entropy fail for the GFF. For example, the $d=2$ entropic c-theorem \cite{Casini:2004bw} requires $(r\,S'(r))'\le 0$. This is clearly not the case for and entropy growing as the size of the interval $S\propto \,r$. The reason is clear. The irreversibility theorems are a consequence of causality and Lorentz invariance, combined with strong subadditivity, and the derivation does not hold for non causal GFF.    

We have focused on conformal GFF. Other GFF can be studied holographically as well using asymptotically AdS space-times. For space-times with an IR AdS fix point we still have a volume term, with the same coefficient. However, for a gapped boundary theory the bulk ends at a certain distance from the boundary, and we expect to recover an area term as in the Kaluza Klein models. This points to an IR origin of the volume term in the holographic description, which is at the same time an UV divergent term because of its dependence on $\epsilon$. 

\section*{Acknowledgments}
 This work was partially supported by CONICET, CNEA
and Universidad Nacional de Cuyo, Argentina. The work of H. C. is partially supported by an It From Qubit grant by the Simons foundation.


\bibliographystyle{utphys}
\bibliography{EE}

\end{document}